\documentclass[a4paper,fleqn,usenatbib]{mnras}

\usepackage{newtxtext,newtxmath}

\makeatletter

\newcommand{\Rmnum}[1]{\expandafter\@slowromancap\romannumeral #1@}

\makeatother

\usepackage{graphicx}	
\usepackage{amsmath}	
\usepackage{subcaption}
\usepackage{color}

\def\apj{{\rm ApJ}}
\def\aj{{\rm AJ}}
\def\apjs{{\rm ApJS}}

\def\mnras{{\rm MNRAS}}

\def\kpc{\, {\rm kpc}}
\def\hmpc{h^{-1}{\rm Mpc}}
\def\hgpc{\;h^{-1}{\rm Gpc}}

\def\kms{\, {\rm km}\, {\rm s}^{-1}}

\def\msun{\, M_{\odot}}

\def\simlt{\lower.5ex\hbox{$\; \buildrel < \over \sim \;$}}
\def\simgt{\lower.5ex\hbox{$\; \buildrel > \over \sim \;$}}

\title[Hubble constant from LSST/NGRST galaxy parallax]{
  Direct geometrical
  measurement of the Hubble constant from galaxy parallax: predictions for
  the Vera C. Rubin Observatory and Nancy Grace Roman Space Telescope
}

\author[R.A.C. Croft]{\parbox{18cm}{
Rupert A.C. Croft$^{1,2,3}$\thanks{E-mail: rcroft@cmu.edu}
}\vspace{0.3cm}\\
$^{1}$ McWilliams Center for Cosmology, Dept. of Physics, 
Carnegie   Mellon  University, Pittsburgh, PA 15213, USA\\
$^{2}$ School of Physics, The University of Melbourne, VIC 3010, Australia\\
$^{3}$ ARC Centre of Excellence for All Sky Astrophysics in 3 Dimensions (ASTRO 3D), Australia\\
}

\begin{document}

\topmargin=-1.0cm

\maketitle


\begin{abstract}

We investigate the possibility that a statistical detection of the
galaxy parallax shifts
due to the Earth's motion with respect to the CMB frame (cosmic secular parallax)
could be made by the Vera C. Rubin Observatory Legacy Survey of Space and Time (LSST) or by the Nancy Grace Roman Space Telescope (NGRST),
and used to measure the Hubble constant. We make mock galaxy surveys which
extend to redshift $z=0.06$ from a large
N-body simulation, and include astrometric errors from the LSST  and NGRST science
requirements, redshift errors and peculiar velocities. We include 
 spectroscopic redshifts for the brightest
galaxies ($r < 18$) in the fiducial case. We
use these catalogues to make measurements of parallax versus redshift,
for various assumed survey parameters and analysis techniques. We find
that in order to make a competitive measurement it will be necessary
to model and correct for
the peculiar velocity component of galaxy proper motions. It will also be necessary to push astrometry
of extended sources into a new regime, and 
combine information from the different elements of resolved galaxies. In an appendix we describe some simple tests of galaxy image registration which yield relatively promising results.
For our fiducial
survey parameters, we predict an {\it rms} error on the direct geometrical measurement of  $H_{0}$ of $2.8\%$ for LSST and $0.8\%$ for
NGRST.

\end{abstract}


\section{Introduction}
\label{intro}

There has been much recent evidence for  tension between different measurements of
the Hubble constant \citep[e.g.,][]{freedman2019,riess2019}. Discrepancies between
local and early Universe measurements \citep{verde2019}
have been interpreted by some as evidence for
 problems with the standard cosmological model. On the other hand, it is indisputably difficult to measure cosmological distances, as shown by
the history of Hubble constant
measurements \citep[e.g.,][]{huchra92, croft11, crossland20}. It is clear
that new techniques would be helpful, and in principle the more direct the
better. Most basic would be to use geometry: parallax measurements, but
so far these have  only been used for distances within our galaxy. The baseline for annual parallax
is a limiting factor in this case. An alternative for observing parallax shifts of objects 
 beyond our galaxy is the continually increasing baseline caused by
the Earth's motion with respect to CMB frame \citep[][ hereafter DC]{kardashev73,ding2009}. 
This cosmic secular parallax could be detected by the  Gaia
satellite (DC, \citealt{paine20,hall19}), but the LSST offers a new 
opportunity, assuming that considerable difficulties with ground based astrometry
can be overcome. Here we explore what might be possible in the best case scenario. We also explore what a future space based measurement with NGRST could offer.

There are many different 
methods used currently to measure the Hubble Constant,
including supernovae and cepheids as  standard candles \citep[e.g.,][]{dhawan18,riess2005,riess2019,freedman01,hubble1925}, gravitational
wave (GW) standard sirens \citep[e.g.,][]{holz2005}, gravitational lens (GL) time delay \citep[e.g.,][]{refsdal1966,chen2019}, and
Baryon Acoustic Oscillations (BAO) in the CDM model as a standard ruler \citep[e.g.,][]{cuceu19,beutler11}.
GW and GL probes are direct, without need for absolute calibration, but the
modeling involved can be complex. The SN and Cepheids are rungs on the cosmic
distance ladder, but not at the bottom, and 
so  still need parallax measurements. So far, these annual parallax measurements are readily
measurable out to $\sim 10 \kpc$ \citep{gaiadr2,lindegren2018,bailer2018}. To do better with space based observations 
 new satellites have been proposed \citep[e.g.,][]{theia}. NGRST (see below) and Euclid \citep{sanderson17,euclid2019}, which are more general
telescopes  will also have great astrometry potential. 

The Nancy Grace Roman Space Telescope (NGRST, formerly known as WFIRST \citealt{spergel13}) is a NASA observatory
designed to use wide field imaging and slitless
spectroscopy to study a range of topics including dark energy, exoplanets and infrared astrophysics. The satellite's Wide-Field Instrument (WFI) provides a sharp point spread function, precision photometry, and stable observations. The field of 
view is 0.28 square degrees, allowing efficient coverage of large regions of the sky. A primary mission lifetime of 5 years (with a possible extension to 10) will provide a baseline
for measurement of proper motions, and
the dataset promises to be a superb astrometric resource. The currently planned launch date is 2025.

The Vera Rubin Observatory is an almost completed ground based observatory\citep{lsst02,tyson2003} which will survey the entire southern
hemisphere every few nights for 10 years \citep{marshall17} to carry out the Legacy Survey of Space and Time (LSST). While ground based, the LSST is planned to have excellent photometry with unsurpassed
time domain qualities, 
 relevant to $H_{0}$ measurement (e.g., for variable stars,
SN,  the quasar variability needed for GL time delay). The astrometry carried
out by the telescope will also lead to an enormous dataset, with 
billions of objects measured \citep{graham2019,sciencebook}. For the purposes of this
paper, for both NGRST and the LSST, we are most concerned with the  relative astrometry between 
distant quasars/galaxies and those within  a few hundred Megaparsecs. This is because
we are interested in cosmic secular parallax.

Cosmic parallax \citep{kardashev73,ding2009} is one of the possible observables in
the field of "real time cosmology" \citep{quercellini2012,darling2012,darling2018,kor18}. Other examples 
are  the redshift drift \citep{sandage62,loeb1998}, the real time change in CMB anisotropies
\citep{lange07}, or the CMB temperature \citep{abitbol}. One
could imagine others e.g., a time varying Tolman test \citep{tolman34} etc., but it
is clear that many ideas are futuristic at best. Neverthless, there appears to
be an interesting path forward for some
of these observables. 
Pioneering work to set current 
limits has been carried out by
\cite{darling2012}, for
redshift drift from 21cm radio observations, 
and \cite{paine20} for cosmic parallax
from Gaia.

Large new facilities (e.g., Rubin Observatory, NGRST)
can be used to carry out precision cosmology using well tested probes
\citep[e.g.,][]{chisari2019} but also may be able
to detect more futuristic effects such as those from real time cosmology. The latter
may perhaps
even be carried out with the accuracy required to make competitive constraints.
They may also yield surprises, which could be their most interesting aspect.

Our plan for this paper is as follows. In Section \ref{secparallax}, we review the concept
of cosmic secular parallax. In Section \ref{mocks} we describe our mock Rubin Observatory
LSST and NGRST surveys,
including a brief introduction to the Rubin Observatory and NGRST, the cosmological Nbody simulations we use,
and how we include various observational effects in the mocks. In Section \ref{method} we
describe how we we measure the secular parallax as a function of redshift
from the mock surveys. In Section \ref{results} we explain how 
we determine the estimated error on $H_{0}$ measurements, and show how the error depends
on survey parameters and variations in the analysis techniques. In Section \ref{summary} we summarise our conclusions and discuss the possible ways forward. In Appendix A, we
describe some simple test  measurements of galaxy image shifts using observational data.

\section{Cosmic parallax}
\label{secparallax}
Unlike the annular parallax, which repeats at a
constant (small) value, the Earth's motion with respect to the CMB
provides a much longer usable baseline for parallax measurements.
This latter effect is a variant of the ``secular'' 
parallax  \citep[see e.g.,][Section 2.2.3]{binney98}, and is in
 principle
easier to measure because the signal increases linearly with time. \cite{kardashev73} were
the first to propose the cosmic version of the
secular parallax as a means for measuring 
distances outside our galaxy. DC revisited the
idea of this cosmic parallax in the era of modern
CMB measurements and the CDM cosmological model. \cite{paine20} were the first to use observational data to try to measure the effect, using Gaia Data Release 2 to find an upper limit, although this was still a
factor of $\sim40$ above the prediction (see below).

The parallax distance to an object in an expanding Universe was first 
calculated theoretically in the context of
the usual annual parallax and published by \cite{mccrea1935}. Not suprisingly,  he noted
that it was unlikely to be measurable.
 It appears also in the
textbook by \cite{weinberg1972}.
The solar system is moving with respect to the CMB frame
at a velocity of $369 \pm 0.9 \kms$ towards an apex 
with galactic latitude and longitude $l=263.99^{\circ} \pm 0.14^{\circ},
b=48.26^{\circ} \pm 0.05^{\circ}$ \citep{kogut93,hinshaw09,planck20}
As a result,
all
extragalactic objects will experience a parallax shift,
increasing linearly with time, towards the
antapex with amplitude proportional to $\sin\beta$,
where $\beta$ is the angle between the object and the direction of
the apex. Over the observing 
period of the LSST (or extended NGRST mission),
a baseline constantly increasing to $l=3800\mu \rm{pc}$ 
after ten years is therefore available
for measures of parallax. This leads to a maximum apparent proper motion of 77.8 $\mu$ arcsec yr$^{-1}$ for galaxies at a 90 deg. angle to  the apex.
We summarize below the expressions for the parallax
shift of a distant extragalactic source (first computed by \citealt{mccrea1935}),
using the notation due to \cite{kardashev73} and \cite{hogg99}.

The equation for the parallax angle $\theta$ is given by:
\begin{equation}
\theta=\frac{\ell}{r}
\label{pieq1}
\end{equation}
where $\ell$ is the baseline. Here $r$ is the comoving
parallax distance, dependent on  
the curvature of the Universe in the following manner:
\begin{equation}
r=
\left\{ \begin{array}{ll}
        \frac{c}{H_0} \frac{1}{\sqrt{\Omega_k}} 
\tanh(\frac{\sqrt{\Omega_k}H_0}{c}D) & \Omega_k>0\\
        D & \Omega_k=0\\
        \frac{c}{H_0} \frac{1}{\sqrt{|\Omega_k|}} 
\tan(\frac{\sqrt{|\Omega_k|}H_0}{c}D) & \Omega_k<0
        \end{array},
\right.
\label{pardisteq}
\end{equation}
where
\begin{equation}
D=c\int^{z}_{0}\frac{dz'}{H(z)}.
\end{equation}
Here $\Omega_k$ is the curvature parameter expressed in terms
of a fraction of the critical energy density.
The parallax distance therefore resembles
 the angular diameter distance for a flat Universe,
 except that $r$ is a comoving rather than proper distance. In the present paper, we restrict all our
analysis to extremely nearby galaxies, with $z < 0.06$. For simplicity, in our calculations we therefore assume Euclidean space, with 
$D=\frac{cz}{H_{0}}$, and $H(z)=H_{0}$ for all $z$ in 
our range of interest.

The parallax shift due to the changing baseline from
the motion of the solar system will manifest itself
as a proper motion, which is the observable we are interested in, and is given by 

\begin{equation}
    \lvert \pi \rvert =77.8 \, r^{-1} \lvert \sin{\beta} \rvert \, \mu \, {\rm arcsec\, yr}^{-1} \, {\rm Mpc},
    \label{pieq}
\end{equation}
where the angle $\beta$ is as defined above. 

\section{Mock observations}
\label{mocks}
We use cosmological
Nbody simulations to construct simple mock astrometry surveys of nearby galaxies, adding cosmological parallax
as well as other physical and observational
effects. We choose the Vera Rubin Observatory's primary dataset, LSST and the Roman Space
Telescope HLS as two to simulate. The techniques
are also applicable to other  future surveys such those carried out by
the Euclid satellite. We have not looked at synergy between LSST and HLS in this paper, but this might be something valuable to explore in future work.

\subsection{The Vera C. Rubin Observatory Legacy Survey of Space and Time}
The Vera C. Rubin Observatory\footnote{\tt https://www.lsst.org/ } is an astronomical observatory
which will be carrying out an unprecedented survey of 
18,000 sq. deg. of the southern hemisphere over the period 2022-2032. Approximately 2 million images will be taken over this
10 year period with a 3.2 gigapixel camera, leading to
at least 100 Petabytes of data. The dataset (named the
"Legacy Survey of Space and Time", LSST) will include time domain photometry for at least 25 billion galaxies,
with better than 0.2 arcsecond pixel sampling. Each part of
the sky will be visited approximately 825 times, with a
(5 $\sigma$) magnitude limit in the stacked images
of $r<27.8$. 

The hundreds of images for each astronomical object
will make the LSST an excellent resource for astrometry.
The science requirements for LSST astrometry are detailed in \cite{ivesic18}.  The LSST is targeted to obtain parallax and proper-motion measurements of comparable accuracy to those of Gaia at its faint limit ($r < 20$) and smoothly extend the error versus magnitude curve deeper by about 5 mag. We make use of the projected astrometric errors for LSST in our mock surveys (see Section \ref{astroerrsec} below), but with 
the considerable added assumption that it will be possible
to carry out differential astrometry of extended objects
(resolved galaxies).

\subsection{The Nancy Grace Roman Space Telescope High Latitude Survey}

After the launch of NGRST, a Wide Field Instrument (WFI) High-Latitude Survey (HLS) will be performed, taking up to 2 years of observations. The HLS will cover over 2,200 square degrees with imaging and low-resolution (grism) spectroscopy. The imaging, in four NIR bands ($Y$, $J$, $H$, and $F184$), will reach $J=26.7 AB$ for point sources. The Y-band magnitude limit (which, for comparison is closest in wavelength to LSST $r$ band) will be 25.8. The slitless spectroscopy will measure redshifts for over 15 million sources at redshift 1.1 to 2.8. Imaging and spectroscopy will support dark energy weak lensing and baryon acoustic oscillation measurements, respectively, and form an invaluable survey for Guest Investigator archival research studies of general astrophysics topics. Astrometry will be one of these topics,
and a detailed exploration of the capabilities
of NGRST in this area is given by \cite{sanderson17}. 

 As with LSST, we are making the assumption in this paper that astrometry of extended objects, namely galaxies, will be possible, and that the well resolved nature of nearby galaxies will allow more accurate measurements to be made than of single point sources. This will require new astrometric techniques, and this is briefly discussed in later sections of the paper.

Because the HLS  covers a smaller fraction of the sky than the LSST, the orientation of the dataset with respect to the CMB dipole direction becomes important. The mean value of $\lvert \sin(\beta) \rvert$ over the NGRST HLS footprint is 0.3, where $\beta$ is angle between pixels and the dipole direction.

\subsection{Simulations}
We use the largest of the publicly available {\ensuremath{\nu}}$^{2}$GC  simulations \citep{ishiyama15,makiya16} to construct mock catalogues.
The run is of a Lambda CDM model (cosmological parameters taken from \citet{planck2014}), in 
a cubical volume $1.120 \hgpc$ on a side, 
with $8192^{3}$ particles. The mass per particle was $2.2 \times 10^{8} \msun$, and \cite{ishiyama15} have constructed
subhalo catalogues, using the subhalo finder of \cite{rockstar}, and which we use. We use the redshift $z=0$ output exclusively. In the entire simulation volume there are $1.04\times10^{9}$ subhalos above a mass limit of $4.4 \times 10^{8} \msun$.

From the single simulation output we excise 27 spheres of radius 175 $\hmpc$. The
centers of the (non overlapping) spheres are set on an equally
spaced grid of size $3\times3\times3$. For the LSST mocks, we assume that the LSST observations are of
half the sky, so we use the top and bottom of each sphere separately, making 54
mock catalogues in all. These mock surveys will not be quite independent, being drawn from
the same volume, but the large scale velocity fields will be drawn from a volume
with large scale density fluctuations.

For the NGRST HLS mocks, we take the HLS footprint to be the area of sky within 26.7 degrees of either the North or South galactic pole, again using each hemisphere excised from the simulation to make one mock survey. The galactic poles are 48 degrees from the CMB apex or antapex, and the area of the spherical cap which makes up each mock HLS is 2200 square degrees.

In each of the 54 hemispheres (LSST coverage) there are $8.3 \times 10^{6}$ subhalos on average, and
in each NGRST HLS there are $9\times 10^{5}$. Only a small fraction in each survey however, being above the magnitude limit for our spectroscopic sample will be used, as described in more detail in Section \ref{spectrosample})
To keep the simulation simple, we assign 
galaxy luminosities to the subhalos using a constant universal mass to $r$-band light ratio of 
300$h$ \citep{bahcall14}, where the $r$ band (centered on 6500\AA) is directly relevant to the LSST. We make the assumption that this holds also at the lowest wavelength filter for the NGRST HLS, the $Y$ band (9000 \AA).
Throughout the paper we do not differentiate between LSST $r$ band and NGRST $Y$ band. We leave
this to future, more detailed work, which should also incorporate modeling of the stability 
advantages of working in the near-infrared.

Additionally, because we are dealing with low redshifts
($z \le 0.06$), we use the relationship between apparent magnitude and luminosity
appropriate for Euclidean space.  The faintest apparent magnitudes for galaxies that we use (again for our 
spectroscopic sample) are $r=18$, in our fiducial
sample, and $r=20$ in one our tests using fainter galaxies (see Section \ref{appmaglim}).  The
LSST and NGRST HLS apparent magnitude limits of $r=27.8$ and $Y=25.8$ are therefore
not relevant to these samples.

\subsection{Secular parallax}
For each mock survey, we add the appropriate component of cosmic secular
parallax to the proper motion vector of each galaxy, including the angle from the CMB motion apex or antapex, $\beta$. The parallax $\pi$ is from Equation \ref{pieq} (assuming
the Euclidean $r=cz/H_{0}$, because $z \ll 1$). 
In order to calculate $\pi$, we use a baseline $\ell$
appropriate for 1 year, which is 77.8 AU. The units of 
$\pi$ are therefore those of proper motion, arcsec yr$^{-1}$.

\subsection{Astrometric errors}
\label{astroerrsec}
Astrometric errors are usually specified only for point sources, as galaxy astrometry
is not usually attempted. The more centrally concentrated the source, the more
accurately the centroid position can be calculated. One would therefore expect galaxies
with their extended surface brightnesses to be more difficult to use for astrometry. In
the case of the nearby $z \le 0.06$ objects however, most galaxies will be 
large enough to subtend many resolution elements, containing in principle independent
information. It should therefore be possible to compute the relative proper motions
of galaxies by making use of all galaxy pixels.
How to do this optimally is something
that is beyond the scope of this paper, but this should be studied in depth in the future,
perhaps using resampled and shifted galaxy images.
We note that sub-pixel image shift estimation is a task
often carried out  in
high performance image processing techniques relevant
for applications in remote sensing, medical imaging, surveillance and computer vision.  Frequency based shift
methods using phase correlations (\citealt{kug75,zitova03})
have been
widely used because of their accuracy and simplicity,
and can deal with
shifts due to translation, rotation or scale changes. 
In Appendix A, we carry out some illustrative tests of galaxy image registration using these methods, with example galaxy data taken from HST and SDSS imaging.

Whichever methods are used to compute the relative shifts
between galaxy images, we make the assumption that the different resolved elements of each galaxy can be averaged
over to give the overall
 astrometric proper motion error of a galaxy, $\sigma_{\rm astro}$:
\begin{equation}
\sigma_{\rm astro}=\sigma_{\rm point}(m_{e})/\sqrt{ N_{\rm elem}},
\label{astroerr}
\end{equation}
where $N_{\rm elem}$ is the number of resolution elements. In our (very limited) tests
with real galaxies in Appendix A, we see an {\it rms} error $\propto N_{\rm elem}^{-0.44\pm0.025}$. This can be compared to the ideal value of $N_{\rm elem}^{-0.5}$ in Equation \ref{astroerr}. Given that the results are not extremely different from the ideal value, and that further work may 
yield improvements, we use Equation \ref{astroerr} in the analysis in the rest of the paper.  It should be borne in mind however that systematic errors or a deeper level of realism may yield instead significantly worse results (which would mean
that the average astrometric errors per galaxy would be larger), and so  our choice is 
optimistic. For example, for galaxies which vary in  $N_{\rm elem}$ by a factor of $10^3$, the
value of $\sigma_{\rm astro}$ will be a factor of 1.5 larger if the slope of the power relation with
 $N_{\rm elem}$ is -0.44 rather than -0.5. If the slope is instead -0.25, the errors will be a factor of 5.6 worse.

The number of resolution elements itself is 
given by 
\begin{equation}
N_{\rm elem}=(2 r_{\rm half}/0.5r_{\rm res})^{2}.
\label{nelemeq}
\end{equation}
Here $\sigma_{\rm point}(m_{e})$ is the {\it rms} astrometric proper motion error
for a point source of apparent $r$ magnitude $m_{e}$ (see below), $r_{\rm res}$ is the length in kpc corresponding to the median resolution. In the case of the LSST, we use the 
seeing (0.7 arcsec FWHM \footnote{\tt https://www.lsst.org/science/science\_portfolio}),
and for NGRST we use the image pixel resolution (0.11 arcsec, \citealt{spergel13}).
The quantity $r_{\rm half}$ is the galaxy half light radius in kpc. 

We determine $r_{\rm half}$ for each galaxy
in the simulation using 
the approximate relation
\begin{equation}
    \log_{10}{r_{\rm half}}=-0.198 M - 3.117
\end{equation}
taken from \cite{simard99}, where $M$ is the absolute  magnitude of
the galaxy. The absolute magnitudes $M$ were rest frame $M_{B}$ in the case
of the \cite{simard99} data, which ranged from $z=0-1$, but as we are
restricted to $z\sim0$, we use $M=M_{R}$, which is
 a better match. We also make the simplifying assumption that the light from
the galaxy is distributed evenly between the resolution elements, so that the magnitude of each element, $m_{e}$ in 
equation \ref{astroerr} is
given by 
\begin{equation}
m_{e}=m+2.5 \log_{10} N_{\rm elem},
\end{equation}
where $m$ is the apparent $r$ magnitude of the galaxy.

\begin{figure}
 \includegraphics[width=1.0\columnwidth]{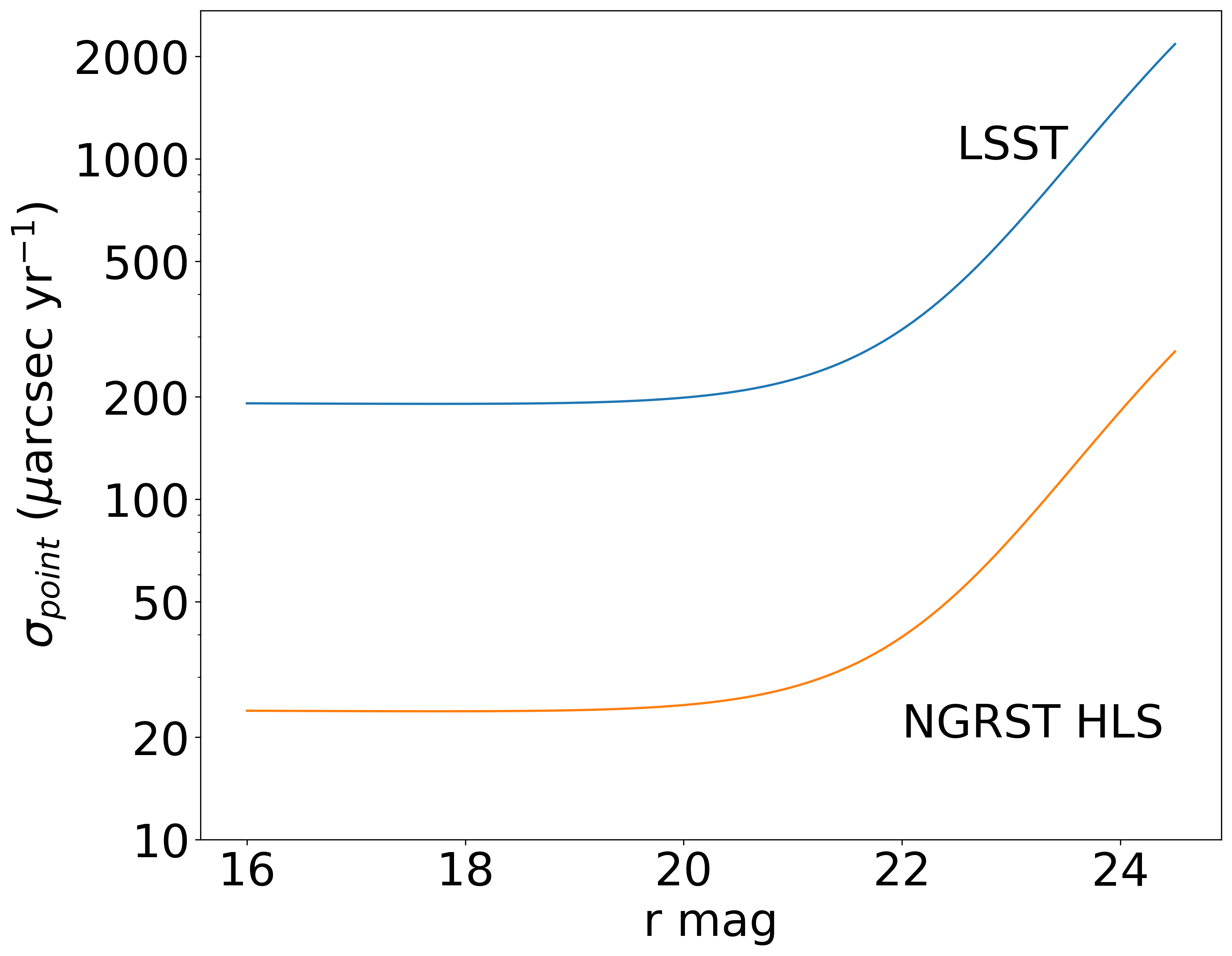}
 \caption{Astrometric proper motion error for a point source as a function of object magnitude, for both
 LSST and NGRST HLS. We show the fits used when generating the mock catalogues (Equation \ref{fitfn}).
}
 \label{errvsmag}
\end{figure}

\subsubsection{LSST astrometric errors}

For the LSST, we take the form of the astrometric error $\sigma_{\rm point}(m)$ for a point source as a function
of magnitude from \cite{ivesic12}, where the relevant quantity is the proper motion error, rather than the parallax error. We use the following fitting function, accurate to within $2\%$:
\begin{multline}
\label{fitfn}
    \log{(\sigma_{\rm point})}=\log{(\sigma_{\rm bright})}+\frac{1}{2}(1+\tanh{(0.522 (m-22.30))})\\
    \times(0.172m-3.80-\log{(\sigma_{\rm bright})},
\end{multline}
where $m$ is the apparent $r$ magnitude, and  $\sigma_{\rm point}$ is in $\mu$arcsec yr$^{-1}$. Here $\sigma_{\rm bright}$ is the constant proper motion error expected
at bright magnitudes.
We plot $\sigma_{\rm point}(m)$ from this function in Figure \ref{errvsmag}, where we can see that it is expected to be approximately constant
until $m\sim20$, before increasing, reaching a maximum of $>2$ milliarcsec at the limiting 
magnitude of the survey. 
The constant error at bright magnitudes quoted by \cite{ivesic12}
is $\sigma_{\rm bright}=\sigma_{\rm point}(m< 20)=130$ $\mu$arcsec yr$^{-1}$, but in Figure \ref{errvsmag} we have used a different value,  to match the LSST Science Requirements,
as follows.

In the LSST Science Requirements document\footnote{\tt https://github.com/lsst-pst/LPM-17},  the design requirement
for astrometric accuracy is relative astrometric precision for a single image of 10 mas (per coordinate).
This translates to a 
a proper motion accuracy of $\sigma_{\rm bright}=200$
$\mu$arcsec yr$^{-1}$. The 
 stretch goal is half of this value. When using 
mock catalogues (below), we treat  $\sigma_{\rm bright}$ as a parameter to 
be varied, using values covering a range from 1 $\mu$arcsec$yr^{-1}$ to 
3000 $\mu$arcsec$yr^{-1}$, with a fiducial value of 200 $\mu$arcsec$yr^{-1}$ (which is
the value shown in Figure \ref{errvsmag}.

In the LSST Science Requirements, the relative {\it rms} astrometric error between two
sources is dependent on  the separation $D$ between them. The requirement above is valid for $D=20$ arcmin, and degrades by 50\% if the relative separation is instead $20 < D < 200$ arcmin.
In our case
we are only interested in the proper motions of a tiny minority of galaxy
sources, those within $z < 0.06 $. Each of these nearby galaxies 
will have its proper motion measured with respect to a frame consisting
of many other more distant
galaxies (and some quasars). For example, we expect that each of the $z < 0.06$
galaxies will have on average $3\times10^{5}$ LSST neighbour galaxies within 20 arcmin (assuming $2\times10^{10}$ LSST galaxies in total). The astrometric error on the frame comprising these galaxies will be suppressed by a large $1/\sqrt{N}$ factor. The frame is therefore fixed by these distant galaxies, and so to a very good approximation the only relevant astrometric error 
for each of the low redshift galaxies is that in Equation \ref{astroerr}. 
 In the mock catalogues, we therefore add the astrometric(proper motion) error to each galaxy, with a direction chosen at random and a magnitude drawn from a Normal distribution with standard deviation $\sigma_{\rm astro}$.

\subsubsection{NGRST astrometric errors}

As a space based telescope, NGRST will continue  the astrometric tradition of the Hipparcos and Gaia satellites. Compared to those instruments, it will have a much larger aperture, 
enabling observations of fainter stars. Crucially for this project,
it is a multipurpose observing instrument, and will make images
of the sky in four near infrared bands with high resolution. 
It is these images that we hope can be used to measure the proper
motions of extended objects. Gaia and Hipparcos were limited to
astrometry of point source objects,  although the bright regions
of galaxies can be registered as point sources (and were used to
set limits on the cosmic parallax by \citealt{paine20} ). 

An overview of the astrometric precision that will be possible with NGRST and a study of the topics which can be addressed is given by \cite{sanderson17}. These include measurements of the motions of stars in the distant Milky Way halo, constraints on dark matter from time dependent quasar lensing, and detection and characterisation of 
exoplanets. \cite{sanderson17} also address the detection of
proper motions of local group galaxies, extending the reach of HST and Gaia by looking at individual stars such as bright K giants. The use of more distant galaxies to measure proper motions, such as we propose, it much more speculative, and would have to rely on techniques for dealing with extended images which have not been applied in the context of astrometry (see e.g., Section \ref{astroerrsec} above ). Although the apparent
proper motions  caused by cosmic parallax are miniscule on an individual basis, the wide area coverage of the HLS will allow thousands of well resolved galaxies within
$z=0.06$ to be used to make a statistical measurement. As with the LSST case, the
relative proper motions of nearby galaxies with respect to a larger number of more distant objects within 20 arcmins or less will be the quantity to measure. It is to 
be hoped that the requirement for relative astrometry rather than an absolute
astrometric solution over the whole HLS footprint will be more forgiving.

\cite{sanderson17} also carried out a comprehensive analysis of the various physical effects which will need to be dealt with to achieve high precision astrometry. These include geometric distortion, pixel level (e.g., quantum efficiency) effects, colour dependence, and readout hysteresis. How best to schedule observations to maximise the astrometric value of various surveys (for example, how to organise the visits to the
HLS fields over the 5 year nominal lifetime of NGRST) is also addressed. These issues will also be relevant for the extended object photometry that we are interested in here, and indeed there are likely to be additional concerns. As for the case
of LSST, we do not address them here, but instead for the purposes of
our simplified calculations assume that the point source 
proper motion precision can be applied to resolved elements of individual galaxies.
\cite{sanderson17} list the approximate astrometric performance of the NGRST Wide-Field Imager, and for relative proper motions derived from  the High-Latitude Survey,
{\it rms} errors of 25 $\mu$ arcsec yr$^{-1}$ are expected for sources well above the magnitude limit. To set the proper motion
error as a function of magnitude in our mock surveys, 
we set  $\sigma_{\rm bright}=25 \mu$ arcsec yr$^{-1}$ in Equation \ref{fitfn}. We therefore use the same fitting function as for LSST, but with a smaller value for this 
parameter. The expected behaviour of the curve at faint magnitudes has not been estimated for
NGRST, but the magnitude limit of the HLS is similar to LSST, and so with the advantages of space-based observations we feel that it is relatively conservative to
assume the same form given by Equation \ref{fitfn} for NGRST.

\subsection{Photometric redshifts}
To constrain the redshift-distance relation, we need a parallax measurement that is a function of redshift. Unfortunately measuring
spectroscopic redshifts for all galaxies observed by the LSST is unlikely to be possible 
in anywhere near the lifetime of the survey (we address the NGRST case below). 
In tests, we have explored the use of photometric redshifts exclusively, but have found
that for the nearby galaxies ($z<0.06$) relevant for our present study, the error in the
inferred distance is too large to allow the use of a velocity flow model correction (which 
is essential, see below). We therefore envisage the use of a subsample of galaxies with
spectroscopic redshifts (to bin the parallax measurement vs. redshift). The rest
of the galaxies with photometric redshifts will be used to define a distant frame which
is at rest (as described above).

For photometric redshifts therefore, we only require that they be precise enough to
differentiate between the nearby galaxies in our spectroscopic sample and those in the
distant sample. Photometric redshifts will themselves be useful for targeting
of the spectroscopic sample, in order to apply a volume limit.
Predictions for and simulated tests of methods for the measurement of
photometric redshifts for LSST galaxies have been made by \cite{graham18}. The 
precision achieved in tests of the situation after 10 years of observations (over the redshift range $z=0.3-3.0$) is  $\sigma_{\rm z}/(1+z)=0.0165 \pm 0.001$, with a fraction of outliers equal to $0.04$.  

Although the NGRST wide field camera operates in four photometric bands in the infrared, it will not be necessary to derive photometric redshifts from them, due to the grism observations which will also be available. The grism \cite{spergel13}
will have $R=700$, which may be sufficient to achieve the redshift accuracy
needed for the project, or else spectroscopic redshifts for the galaxy sample
required could be obtained from another source.




\subsection{Spectroscopic redshifts}
\label{spectrosample}
Because we will be binning the measured cosmic parallax as a function
of redshift from the observer, it is necessary to have an observed 
redshift associated with the galaxies in the mock surveys. We
use the low $z$ approximation, $z=H_{0}d/c+v_{\rm pec}/c$ to assign   redshifts $z$ to each galaxy, where $v_{\rm pec}$ is the peculiar velocity component along the line of sight to the observer. We then
assume that spectroscopic redshifts are available
(for both LSST and NGRST HLS)
for a small
fraction of observed galaxies above a certain
apparent magnitude limit. This fraction is varied in Section \ref{appmaglim}, but we use $r < 18$ to define it in the fiducial case. We add also
a measurement error taken from a  Normal distribution with 
 $\sigma=100$ $\kms$ to each galaxy's redshift. 
 
 In the case of the LSST, this accuracy
 could be achieved from a separate redshift survey with comparable redshift 
 precision to e.g., the SDSS  \citep{blanton05}, but in the southern
 hemisphere. For NGRST HLS, the sky area and therefore number of galaxies with
 $r<18$ is much smaller, and so it should not be difficult to obtain redshifts
 for them even if the NGRST grism itself is unsuitable. As we see will
 see in Section \ref{appmaglim}, the number of redshifts required below this
 magnitude limit for NGRST would be about $40,000$.

\subsection{Peculiar velocities}
The three-dimensional {\it rms} peculiar velocity of galaxies in the simulation is
$\sigma_{V3D}=  550 \kms$. This will lead to both shifts in the Hubble
redshift diagram as well as a peculiar velocity error component (which could be deemed an actual  proper
motion) to the measured
parallax. On a galaxy by galaxy basis, this proper motion component is relatively
small. For example, a galaxy with redshift $z$ in our mock survey will have an
{\it rms} proper motion due to peculiar velocities of $0.8 \frac{0.05}{z} \mu$arcsec yr$^{-1}$.
This would seem to be much smaller than the astrometric error $\sigma_{\rm astro}$ 
for all except extremely nearby galaxies (for example galaxies
within 1$\hmpc$ will have {\it rms} proper motions of $\sim100\mu$arcsec yr$^{-1}$). Unfortunately, the large scale velocity field
has a high degree of coherence \citep[e.g.,][]{gorski89}, and as a result these proper motion errors
do not decrease by simple $\sqrt{N}$ averaging. Indeed, as we shall see below, lack
of knowledge of the
peculiar velocity field over the survey volume will be the main limiting factor for
the precision of the cosmic parallax measurement.

The effect of peculiar velocities has of course been noted by many authors 
(e.g., \citealt{howlett20,mukh19,nico19}
in the context of distorting the redshift-distance relation for other
probes, such as standard sirens. In these studies, corrections have been applied using 
models for the peculiar velocity flow field derived from the gravity
field  of large-scale structure (such as \citealt{springob14}).
\cite{paine20} have extended this analysis to cosmic parallax, where
the predicted angular proper motion of galaxies was examined based on
the Cosmicflows-3 galaxy peculiar velocity catalogue (\citealt{tully16}). 
 In our case, we will examine the effect of peculiar velocities
 in our simulated mock surveys. We will study
 cases where the velocity field is
not corrected, and also where predictions for or measurements of the peculiar velocities are used
to correct them to a certain degree of accuracy.

In our mock catalogues we therefore add the component
of the peculiar velocities of each galaxy along the observer's line of sight to the 
measured redshift. We also add the components of the proper motion due to the
peculiar velocity to the measured proper motion.

\section{Determination of the Hubble constant: method}
\label{method}

\subsection{Measurement of galaxy parallax}
\label{measurement}
For a mock (or real survey), we have information on the measured total astrometric proper motion of each galaxy, which includes cosmic parallax,
measurement errors and peculiar velocity component, as well as the spectroscopic redshift of
each galaxy, again including redshift errors and peculiar velocities. To analyze each mock
and measure the Hubble constant from it, we first divide the galaxies into
bins of measured spectroscopic redshift, and for each bin compute the weighted mean of the proper motion
component in the direction expected for cosmic parallax. The measured parallax is then

\begin{equation}
    \pi(z)= \frac{1}{\sum_{i=1}^{N(z)}w_{i}}                            
\sum^{N(z)}_{i=1} w_{i}\, \sqrt{1-(\hat{n}\cdot \vec{\mu_{i}})^{2}}, 
\label{measurementeq}
\end{equation}
where the sum $i$ is over the $N(z)$ galaxies, in the bin centered on redshift $z$. 
Here  $\hat{n}$ is a unit vector on the surface of the celestial sphere pointing towards the apex of the CMB motion, and $\vec{\mu_{i}}$ is a vector representing the measured astrometric proper motion of galaxy $i$.
The inverse variance weighting 
used is 
\begin{equation}
    w_{i}=\frac{1}{\sigma^{2}_{\rm pm}+\sigma^{2}_{\rm astro,i}},
    \label{weighteq}
\end{equation}
where $\sigma^{2}_{\rm astro,i}$ is the square of the {\it rms} astrometric measurement error for galaxy $i$ (from Equation \ref{astroerr}).  
Here, $\sigma^{2}_{\rm pm}$ is the square of the {\it rms} proper motion for galaxies at redshift $z$, computed
from $\sigma_{\rm pm}=\sigma_{v}/r(z)$ where $\sigma_{v}$ is the {\it rms} peculiar velocity, and $r(z)$ is the parallax distance to redshift $z$, from Equation \ref{pardisteq}. As described in Section \ref{astroerrsec}, in our fiducial case, we make use of the separate elements of each resolved galaxy
to determine $\sigma^{2}_{\rm astro,i}$ for the galaxy.

We compute the error bars on $\pi(z)$ from the standard deviation of the values from all 54 mock
catalogues, and also compute the covariance matrix between redshift bins. In order to determine the value
of the Hubble constant, we compare the measured $\pi(z)$ to a theoretical curve and fit an
amplitude parameter. The theoretical curve is derived from Equation \ref{pieq}, but smoothed
by peculiar velocities and redshift errors. 
These would have a substantial impact if only photometric redshifts were
used, but in our case, as spectroscopic redshifts are employed to bin galaxies.
so the difference between  Equation \ref{pieq} and the smoothed version is 
small, as we see below.

\subsection{Predicted parallax}
\label{acczerrs}
In order to estimate the Hubble constant, it is necessary to compare
the results of a cosmic parallax measurement (from Equation \ref{measurementeq})
with a prediction. The predicted parallax for a set of objects with no peculiar velocities, and redshifts which fall exactly at the centers of 
the redshift bins used in the measurement, and which are all at the same angle with respect to the  apex of the CMB motion can be computed using 
Equation \ref{pieq}. At the level of accuracy required to make competitive estimates
of the Hubble constant however, the non uniformity in redshift and angle of the
sample of galaxies used will result in any binned estimate of redshift, angle, and 
parallax having significant Poisson errors. We therefore include information which will be available from observations
for the individual galaxies when making the prediction. This information is their
angular positions, their spectroscopic redshift, and when relevant, 
estimates of their peculiar velocities from a flow model (see Section \ref{flowmodel}).
The predicted parallax for an assumed value of $H_{0}$ is therefore:

\begin{equation}
    \pi_{\rm p}(z)= \frac{1}{\sum_{i=1}^{N_{p}(z)}w_{i}}                            
\sum^{N_{p}(z)}_{i=1} w_{i}\, \pi(H_{0},z_{p,i}) \left|\sin{\theta_{i}}\right| 
\label{predictionpi}
\end{equation}

where the sum $i$ is over the $N_{p}(z)$ galaxies, in the bin centered on 
redshift $z$. Here $\theta_{i}$ is the angle between the position of galaxy  $i$ and the apex due to the CMB motion. Equation \ref{pieq} is used to compute
$\pi(H_{0},z_{p,i})$, where $z_{p,i}$ is
the spectroscopic redshift of galaxy $i$. If a
flow field model (Section \ref{flowmodel}) is used,
$z_{p,i}$ is instead the spectroscopic redshift minus the line of sight component of the flow field peculiar  velocity. The weight $w_{i}$ is taken
from Equation \ref{weighteq}.

\subsection{Use of spectroscopic redshifts}

Because we are focusing on low redshift observations $z<0.06$, we find that the LSST photometric redshift accuracy $\sigma_{z}\sim0.02$ \citep{graham2019} is too low to allow
precise measurement of $\pi(z)$. This is not because of the smoothing effect
of redshift errors on the $\pi(z)$ curve, but instead because the photometric
redshifts do not allow the galaxy three dimensional positions to be estimated
well enough to apply a peculiar velocity correction (see below). In order to
do this, we estimate that a sample of spectroscopic redshifts (for 
both LSST and NGRST HLS will need to 
be collected, for the brightest galaxies within a photometric redshift limit . 

For our fiducial analysis of mocks, we assume that spectroscopic redshifts will
be available for galaxies with $r<18$. This is similar to the SDSS main galaxy sample, which had Petrosian $r <17.77$. The use of photometric redshifts to preselect low redshift targets will further reduce the total number of redshift measurements needed. At present the largest planned samples of
galaxies with spectrocopic redshifts are in the northern hemisphere (DESI, \citealt{schlegel15}, and
WEAVE, \citealt{weave16} ), but the redshift requirement for the cosmic parallax measurement is  modest compared to these surveys. For example, we find that in our
mock surveys that if all galaxies with photometric redshift $z<0.06$ and $r<18$
are targeted, the sample per survey is about 400,000 galaxies for LSST and 40,000 for the HLS. In our analysis we will vary the magnitude limit, and find that competitive estimates of $H_0$ could be possible even with a magnitude limit as low as $r<14$ for spectroscopic redshifts, which would be only 12,000 galaxies on average from an LSST survey.

The photometric redshifts of all galaxies are still useful for the parallax
measurement. Even though we only directly use galaxies with spectroscopic
redshifts in our binned $\pi(z)$ determination, the parallax measurements
for those galaxies will have been obtained as differential measurements, where the angular displacements are obtained with respect to a frame defined by  much more numerous
distant galaxies that are close on the sky. That they are more distant will be ensured through the
use of the photometric redshifts available for all galaxies brighter than
$r=27.8$ for LSST, and $Y=25.8$ for NGRST HLS.

\subsection{Accounting for peculiar velocities}
\label{flowmodel}
Peculiar velocities will distort the parallax distance-redshift relation. Because we concern ourselves here with nearby galaxies, with recession velocities $cz < 18,000 \kms$, the coherent flows of galaxies due to gravitational instability will be a more important factor than they are for e.g.,
cosmology with Type IA supernovae \citep{davis11}. It is customary to correct galaxy redshift for the effects of peculiar velocities using a flow model when
determining the nearby distance redshift relation (for example see the early work of e.g., \citealt{willick01} for cepheids, or more recently  \citealt{mukh19}). Flow models
can be constructed using the predicted gravitational potential computed from the galaxy density field (using a bias model) and linear theory or other 
models to
relate the peculiar velocities to the acceleration. Examples of such flow models are \cite{branchini99} and \cite{springob14}. 

The accuracy of such flow 
models can be tested to some extent using N-body simulations. For example, the PIZA algorithm of \cite{croft97} can recover the velocity field
smoothed on scales of 5 $\hmpc$ with an {\it rms} error of $37 \kms$. In general, the velocity field in flow models is predicted on a grid, which 
can be interpolated to the positions of galaxies, leading to a further dispersion of peculiar velocities between galaxies and the interpolated grid. In the present work, we do not apply a particular flow field reconstruction method to our mock catalogues, but instead use the actual smoothed velocity field of
galaxies as a correction, and add an {\it rms} dispersion (a parameter to be varied) to model the effects of non-linearities and galaxy bias. We leave
the implementation and use of an actual flow field reconstruction method (such as \citealt{colombi07}, or \citealt{yu19})
  and tests on simulations to future work.

Our procedure is as follows. We assign the redshift-space positions of the simulated galaxies in the mock surveys to a three dimensional grid of cell size
$5 \hmpc$ using a nearest grid cell assignment scheme. We also assign the peculiar velocities of galaxies to the same grid, weighting by number, to produce a
flow field grid. We smooth the grid with a Gaussian filter with $\sigma=5 \hmpc $. We use this smoothed flow field to correct the galaxy
angular proper motions in the mock survey, by subtracting the flow velocity for the cell containing the galaxy (again in redshift space,
including redshift errors), and adding a random dispersion velocity component. Our fiducial random velocity dispersion is $100 \kms$ (for 
comparison, \citealt{nico19} find a residual velocity error in a galaxy of $36 \kms$, and \citealt{abbott17} find $69 \kms$ ). We also use the flow field
to correct the redshift space positions of galaxies to real space, by subtracting
the line of sight component of flow model velocity from the observed redshift (see e.g., \citealt{gramann94}, \citealt{croft97}, and   \citealt{wang19} for other methods).

When constructing a flow field prediction, the relationship between galaxies and mass needs to be specified, as the mass governs the 
gravitational accelerations. This is an area where the galaxy angular motion information from Rubin Observatory and NGRST will be very useful, and the astrometry be used to validate and calibrate the flow field model.
For example, the linear bias  (the constant of proportionality relating the galaxy and mass overdensities) could
be left as a free parameter. Varying this parameter would change the flow field model
used as a correction. The best fit of $H_{0}$ and the bias parameter together 
could be obtained by comparing to the galaxy angular motion measurements. Measuring
galaxy transverse proper motions
statistically has been discussed by \cite{darling18}, \cite{hall19} and \cite{paine20}. The use of astrometry in this way could make the flow field model and distance estimate of $H_{0}$ internally consistent, and eliminate any need to consider other sources of distance information.

\section{Results}
\label{results}
We have carried out the analysis described above on our mock LSST and NGRST HLS surveys. In order to study how
the results depend on survey analysis parameters, we have varied them from their fiducial values.
These are first described below.

\begin{table}
\begin{tabular}{|l|l|l|l|}
\hline
parameter & description & fiducial value & fiducial value \\ 
 &  & for LSST & for NGRST HLS \\ \hline
$\sigma_{\rm bright}$ & \begin{tabular}[c]{@{}l@{}}$rms$ proper \\ motion error \\ for bright objects\end{tabular} & 200 $\mu$arcsec/yr  & 25 $\mu$arcsec/yr\\ \hline
$r$ mag & \begin{tabular}[c]{@{}l@{}}limiting galaxy\\  $r$ magnitude\end{tabular} & 18 & 18 \\ \hline
$f_{\rm res\ elements}$ & \begin{tabular}[c]{@{}l@{}}fraction of resolved\\  elements used \\ independently\end{tabular} & 0.5 & 0.5\\ \hline
\end{tabular}
\caption{Fiducial parameter values used in mock catalogues. \label{fidvals}}
\end{table}

\begin{figure}
 \includegraphics[width=1.0\columnwidth]{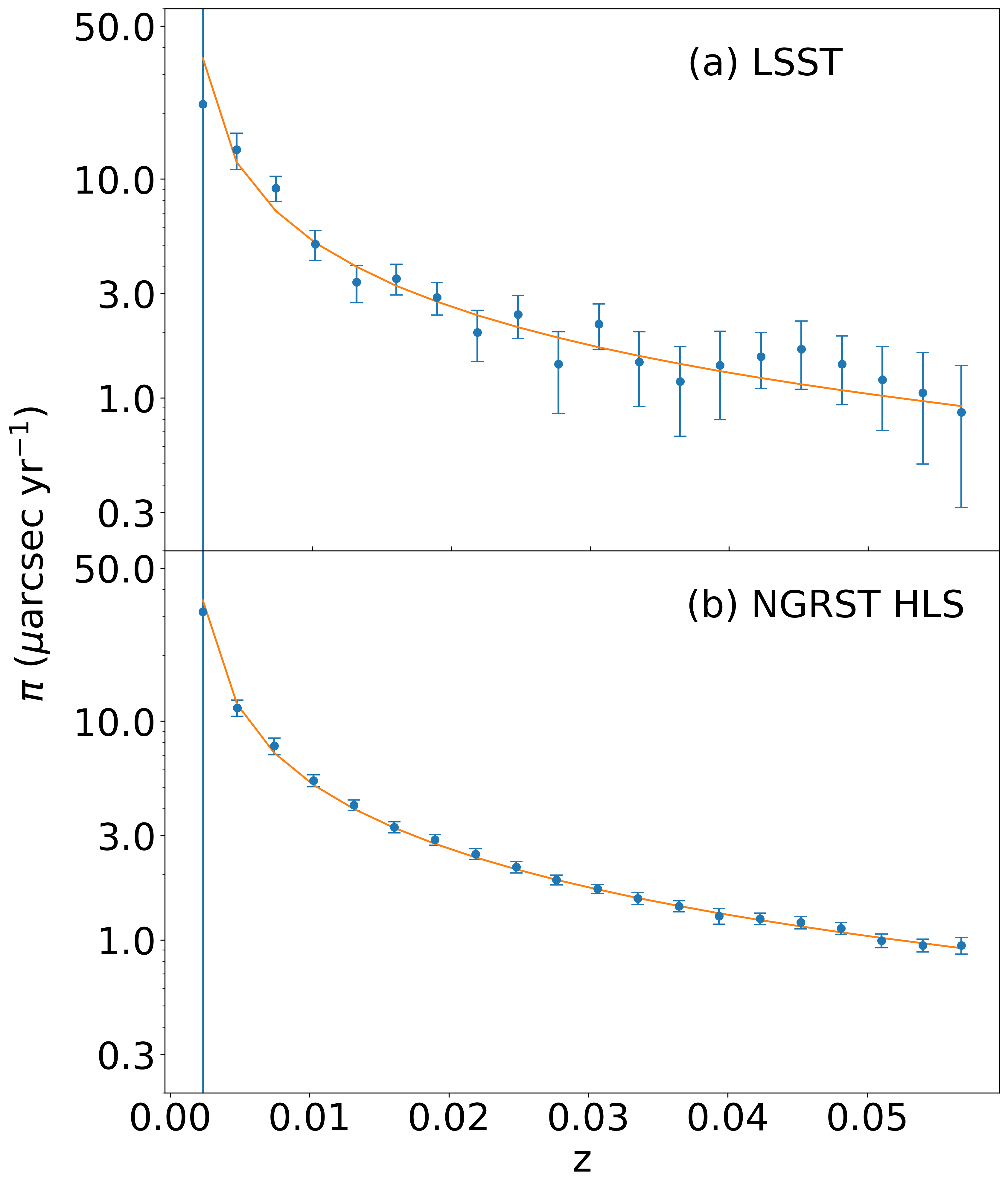}
 \caption{Secular parallax vs. redshift. Points with error bars show results from  (a) an example mock LSST, and (b) a mock NGRST HLS, using  galaxies with $r<18$ (with spectroscopic redshifts). The smooth curve shows the theoretical prediction made using Equation \ref{predictionpi} and with the correct value of
 $H_{0}$.
}
 \label{spectroz}
\end{figure}

\subsection{Fiducial analysis}
\label{fida}
In our fiducial analysis, we assume that the peculiar velocities leading to angular proper
motions (and redshift distortions) can be corrected
with the aid of a flow model (Section \ref{flowmodel}). We investigate what happens
when this is not the case in Section \ref{elems} below. 
Apart from this, we also vary the following three parameters
The first is $\sigma_{\rm bright}$, the {\it rms} proper motion error for bright objects (in Equation \ref{fitfn}), with the fiducial value for LSST being that given
in the LSST science requirements, 200  $\mu$ arcsec yr $^{-1}$, and
the value for NGRST HLS, 25  $\mu$ arcsec yr $^{-1}$ taken from \cite{sanderson17}. The second parameter is
$r$ mag, the limiting $r$ magnitude of galaxies used in the sample (and which have 
spectroscopic redshifts). We choose $r$ mag=18 as the fiducial value. The third parameter 
is $f_{\rm res \, elements}$, and relates the number of resolved elements of a bright galaxy that are used separately to compute the astrometric motions to the total number. We therefore modify Equation \ref{astroerr} to read
\begin{equation}
\sigma_{\rm astro}=\sigma_{\rm point}(m_{e})/(f_{\rm res \, elements} \sqrt{N_{\rm elem}}), 
\label{astroerr2}
\end{equation}
We assume that information from a significant fraction
of the resolution elements would be used, using $f_{\rm res \, elements}=0.5$
as our fiducial value. The true value should be computed with tests on
galaxy images (as discussed  in Section \ref{astroerrsec}). We explore the effect
of widely varying values of $f_{\rm res \, elements}$ in Section \ref{elems}.

 The three parameters and their fiducial values are summarised in Table \ref{fidvals}. With these values, we compute the
 measured parallax as a function of redshift following the methodology in Section \ref{measurement}. 
 We use 20 bins equally spaced in redshift between $z=0$ and $z=0.0583$ (a distance of 175 $\hmpc$).
 We also compute the predicted $\pi(z)$ curve for the correct $H_{0}$ value, in a manner that uses information that would be available from observations (Equation \ref{predictionpi}).
 
 We plot the results for a single randomly
 chosen mock survey in Figure \ref{spectroz}.
 To allow easier comparison between surveys in the plot, we scale both the LSST and NGRST HLS predictions and measured datapoints (and error bars) upwards by 1 divided by the mean of $\sin{\beta}$ over the respective survey footprint. Here $\beta$ is the angle between a galaxy and the CMB apex (Equation \ref{pieq}).
From Figure \ref{spectroz}  we can see the the measurements scatter about the predicted
 line, and that there is measurable signal over the entire redshift range plotted,
 even for LSST, which has the largest error bars. 
 Even at $z\sim 0.06$, the signal to noise per bin is $\sim2$ for LSST and $\sim10$ for 
 NGRST HLS, indicating that it could be productive to extend the measurements to 
 higher redshifts. Because of the relatively small size of our simulated catalogues we leave this to future work.

\begin{figure}
 \includegraphics[width=1.0\columnwidth]{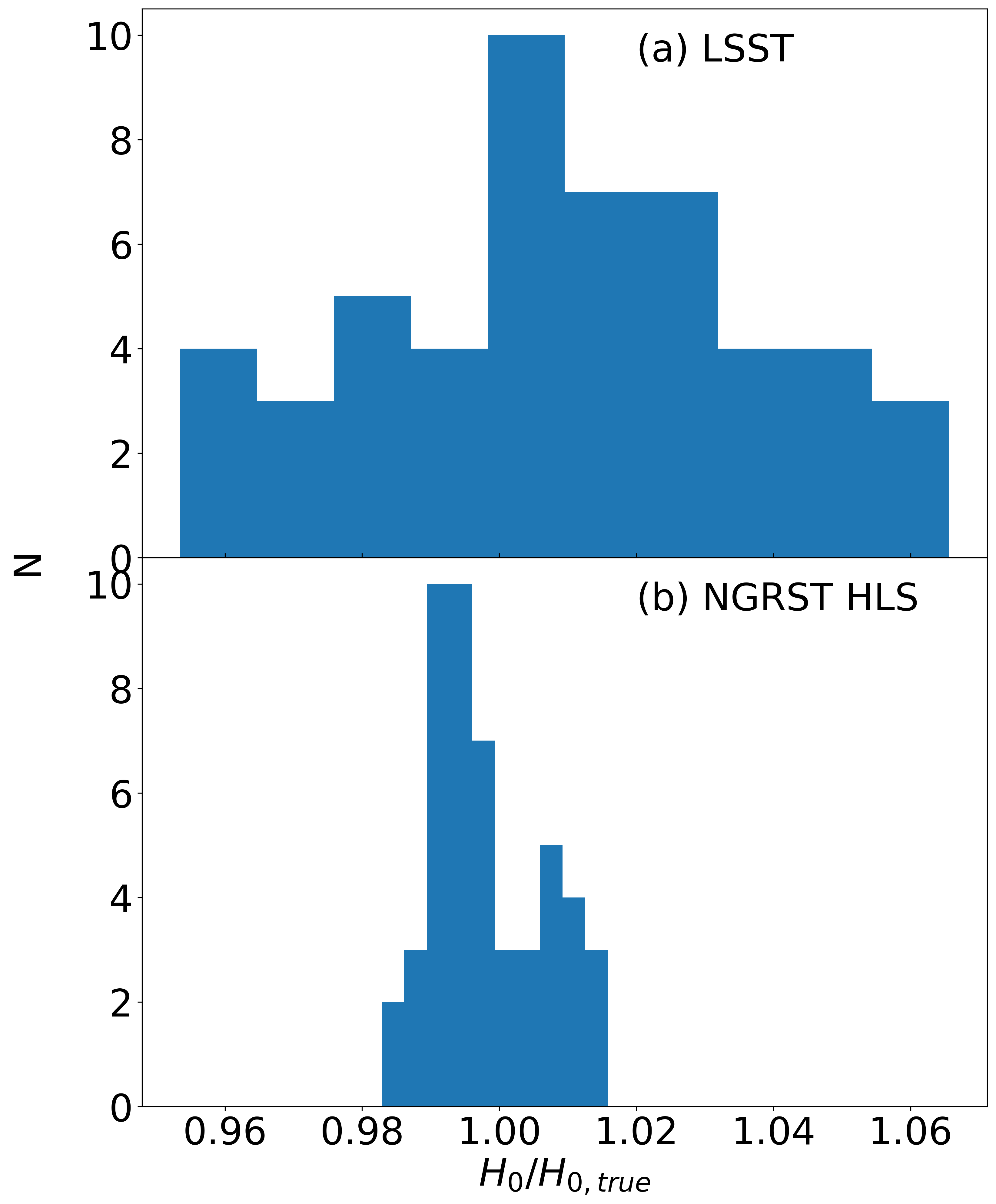}
 \caption{Histograms of $H_{0}$ measurements for (a) mock LSST surveys and (b) mock NGRST HLS surveys. We show results for the fiducial case (Section \ref{fida}).
}
 \label{hhist}
\end{figure}

As stated above, we compute the elements of the covariance matrix of the points in Figure \ref{spectroz} from the scatter between results for the 54 mock catalogues. Then use this covariance matrix to compute the best fitting value of an overall amplitude parameter
scaling the value of $H_{0}$. We use the Python routine  {\texttt{scipy.optimize.curve$\_$fit}} to compute the fit amplitude, using least squares minimization. This yields a best fitting value of $H_{0}/H_{0, true}$ for each catalogue,
as well as a measurement error. A histogram of $H_{0}/H_{0, true}$ for our fiducial
analysis is shown in Figure \ref{hhist}, where we can see that the values lie between
$H_{0}/H_{0, true}=0.95$ and $1.06$ for LSST, being relatively symmetrical,
and $0.98$ and $1.02$ for NGRST HLS.
The mean (median) $H_{0}/H_{0, true}$ values are 1.008 (1.008) for LSST, and 0.998 (0.997) for NGRST.

\begin{figure}
  \includegraphics[width=1.0\columnwidth]{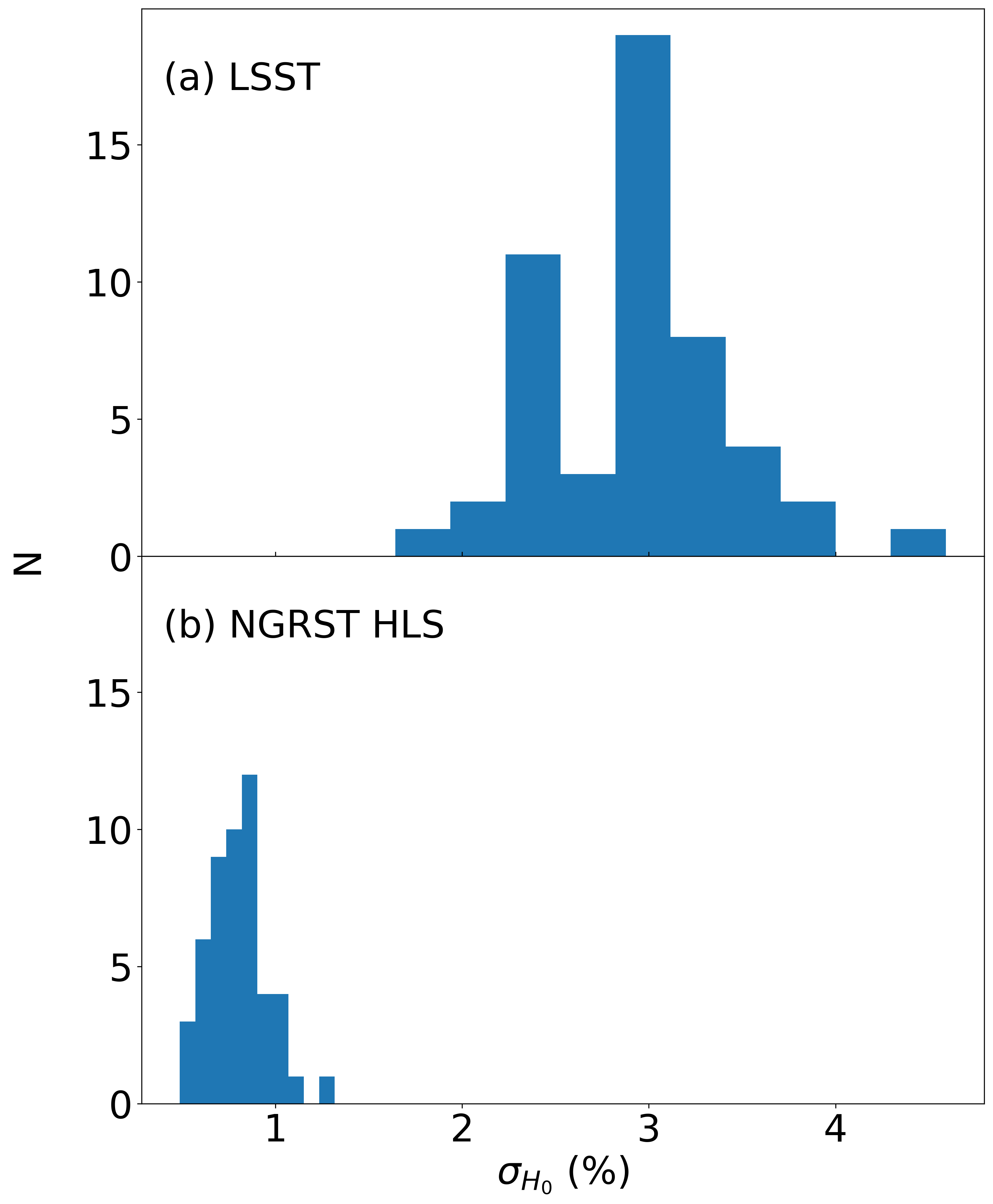}
 \caption{ Histograms of $H_{0}$ fractional measurement errors for (a) mock LSST surveys and (b) mock NGRST HLS surveys. We show results for the fiducial case (Section \ref{fida}).
}
 \label{errhist}
\end{figure}

We compute an estimate of the mean fractional error on $H_{0}$ from the standard deviation
of the results for all mock surveys plotted in Figure \ref{hhist}, finding a value
of 2.8\% for LSST and 0.8\% for NGRST HLS  (these values are for our
fiducial analysis and parameter choices). Error estimates for each mock survey are also available
individually, and a histogram of the one sigma error bar sizes is plotted in Figure \ref{errhist}. 
The mean of these fractional errors is also 2.8\%, with values ranging from 1.6\% to 4.5\% (LSST) and
mean 0.8\% with range 0.5\% to 1.3\% (NGRST). The
$\pm 1 \sigma$ spread of the error estimates is $\pm 0.5$ \% (LSST) and $\pm 0.15\%$ (NGRST). 

Using the error bars on each measurement of $H_{0}/H_{0, true}$ allows us to see
if the measurement is unbiased. Because we have 54 mock surveys, the statistical error on the mean
 $H_{0}/H_{0, true}$ value from all of them is 2.8\%$/\sqrt{54}=$0.38\% for LSST. Our mean
 $H_{0}/H_{0, true}$ ($1.009$, mentioned above) 
 is therefore $(1.008-1.0)/0.0038=2,1 \sigma$ from the expected
 value of $1.0$, and therefore for LSST the method is unbiased at the $\sim 2\sigma$ (based on all 54 surveys). Of course an actual observational measurement would be made from a single survey, and this bias would 
 correspond to $(2.1/\sqrt{54})=0.3$ times the statistical error bar, which is subdominant. This small bias is likely arises because the
 observed redshifts of the galaxies used to make the predictions in Equation \ref{predictionpi} are not the true Hubble redshifts which govern the actual galaxy parallax (Equation \ref{pieq}).  For NGRST HLS, the equivalent calculation also reveals a 2$\sigma$ bias 
 for the mean of all 54 surveys and $0.3 \sigma$ for an individual survey.

\subsection{Impact of astrometric errors}
\label{resultsastroerr}

\begin{figure}
\begin{center}
 \includegraphics[width=1.0\columnwidth]{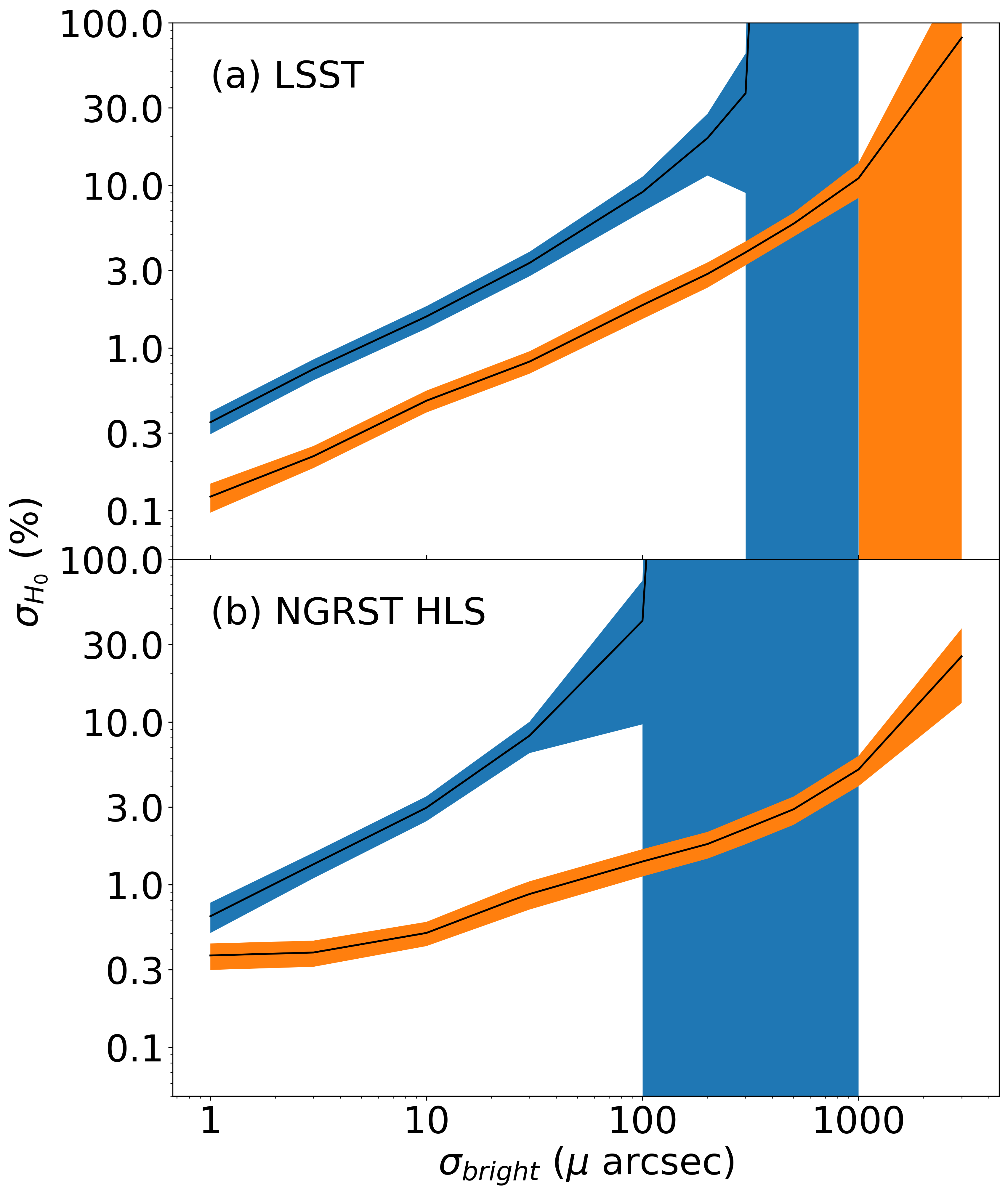}
 \end{center}
 \caption{ The {\it rms} percentage error on $H_{0}$ measurements as a function of the
 astrometric error for bright sources, $\sigma_{\rm bright}$. The results shown in orange are for the case where the rest of the fiducial analysis is kept the same, and the results in blue restrict the number of astrometric measurements to one per galaxy
 (rather than making use of separate resolution elements).
  The envelope around each line illustrates  the standard deviation of the  $\sigma_{H_{0}}$ values for
the mock surveys. The top panel, (a) is for mock LSSTs, and the bottom panel, 
(b) is for mock NGRST HLSs}
 \label{astro_err}
\end{figure}

In our fiducial analysis we used the astrometric error as a function of magnitude from the
 predictions shown in Figure\ref{errvsmag} for LSST and NGRST HLS. Depending on many factors related to
the functioning of the telescopes and the surveys, the astrometric precision achieved could
be better or worse. We parameterise the overall precision using the $\sigma_{\rm bright}$ limiting astrometric error
for bright objects, as explained in Section \ref{fida}. We have varied 
the parameter $\sigma_{\rm bright}$ between 1 and 3000 $\mu$ arcsec yr $^{-1}$, making new
mock surveys for both LSST and NGRST HLS each time, and computing the fractional error on $H_{0}$ measurements, $\sigma_{H_{0}}$. The results are shown in Figure \ref{astro_err}, where the rest of the analysis
is kept as in the fiducial case (orange line). The  $\sigma_{H_{0}}$ values span 0.15\% to 100\% (LSST) and 0.4\% to 25\% (NGRST) over the range of 
$\sigma_{\rm bright}$ values tried.

Our fiducial analysis includes the statistical reduction of astrometric errors caused by the consideration of
the separate resolved elements of bright galaxies. It is interesting to relax this assumption, in order to see
how much worse the distance estimates would be. We do this by degrees in Section \ref{elems} below, but in the present case, we have also plotted in Figure \ref{astro_err} (blue) lines showing results where each galaxy is only used to 
compute one secular parallax measurement, irrespective of size or apparent magnitude. We can see that in this case
the errors on $H_{0}$ are significantly larger, and increase rapidly. For the LSST. when $\sigma_{\rm bright}=10$  $\mu$ arcsec yr $^{-1}$,  $\sigma_{H_{0}}=1.6$\%, increasing to 37\% for  $\sigma_{\rm bright}=300$  $\mu$ arcsec yr. For NGRST, the increase
in  $\sigma_{H_{0}}$ with $\sigma_{\rm bright}$ is even more rapid. This illustrates
that it will be critical to develop techniques to make full use of the resolved elements of large galaxies, and
without this $H_{0}$ measurements are unlikely to be feasible for LSST and likely be uncompetitive for NGRST HLS. We should bear in mind that the samples of galaxies
used in Figure \ref{astro_err} are those brighter than our fiducial apparent magnitude limit of $r=18$, and
are assumed to have spectroscopic redshifts. Therefore
the blue lines make use of many fewer galaxies than will be observed by the LSST or NGRST HLS and will have photometric (or grism) redshifts. Although
the number increase for a photometric sample would decrease the statistical errors, we have found (Section
\ref{flowmodel} that spectroscopic redshifts are essential for correcting the effects of peculiar velocities, at least
for the nearby sample of galaxies considered in this paper.

\subsection{Apparent magnitude limits}
\label{appmaglim}

\begin{figure}
 \includegraphics[width=1.0\columnwidth]{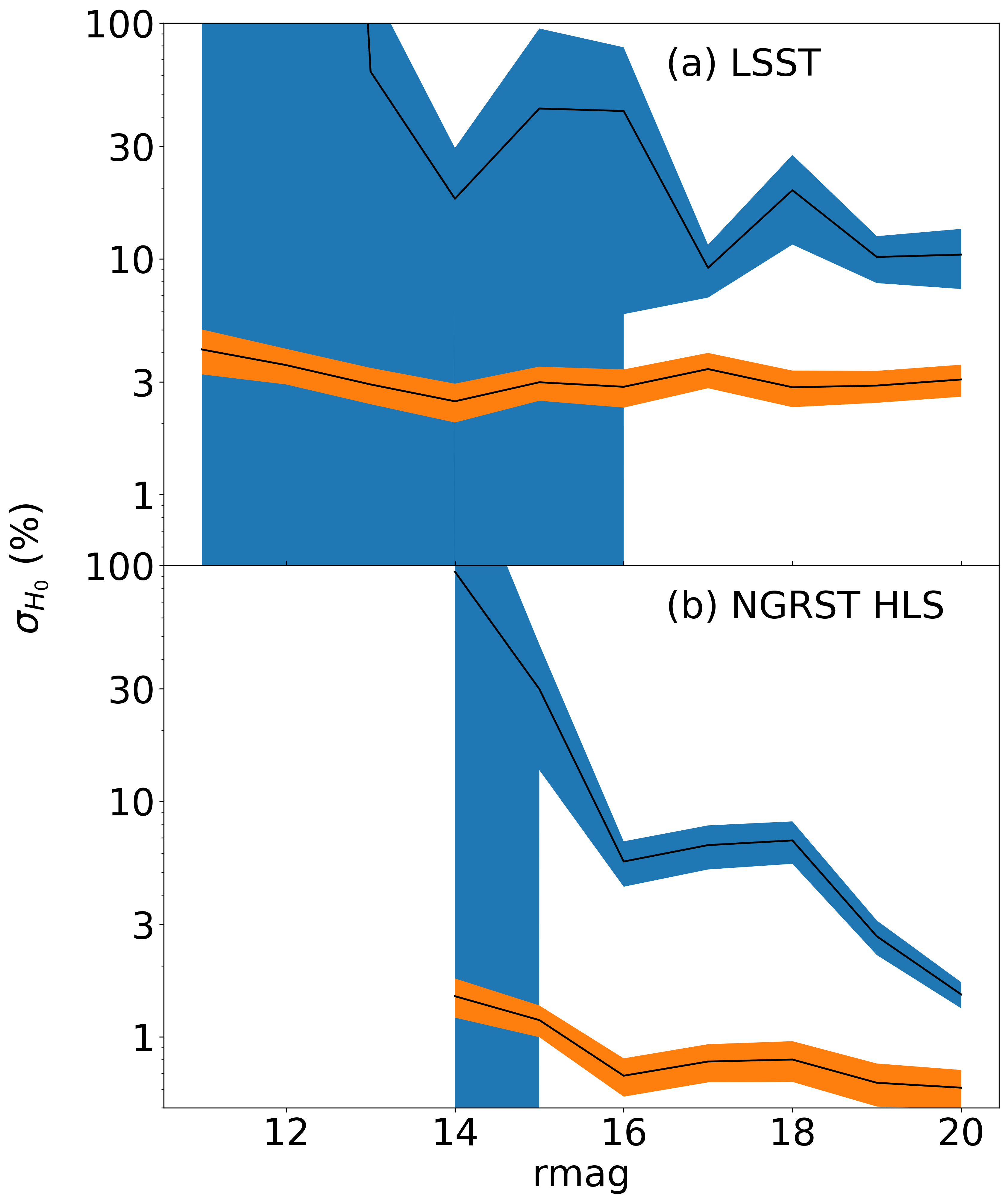}
 \caption{
 The {\it rms} percentage error on $H_{0}$ measurements as a function of magnitude limit
 for galaxies used. The results shown in orange are for the case where the rest of the fiducial analysis is kept the same, and the results in blue restrict the number of astrometric measurements to one per galaxy
 (rather than making use of separate resolution elements).
  The envelope around each line illustrates  the standard deviation of the  $\sigma_{H_{0}}$ values for
the mock surveys. The top panel, (a) is for mock LSSTs, and the bottom panel, 
(b) is for mock NGRST HLSs.
}
 \label{rmaglim}
\end{figure}

We now explore the effect of limiting apparent magnitude on the sample of galaxies used. Our fiducial value is
$r$ mag$=18$, and we assume that only galaxies brighter than this will have spectroscopic redshifts. Fainter
galaxies will still be used to set the astrometric reference frame local to each bright galaxy (see Section \ref{astroerrsec}). When we decrease the limit from $r<18$ to $r<16$, the mean number of galaxies in 
each mock LSST survey decreases from 390,000 to 78,000, but the error on $H_{0}$ only increases from  $\sigma_{H_{0}}=2.7$ \%
to 2.9\%. For NGRST the situation is similar. This is largely a consequence of the fact that the fainter galaxies which are being eliminated make  a
much smaller constribution than the bright galaxies which cover many resolution elements. The yellow curves in 
Figure \ref{rmaglim} show these results,  where we can see that the mean error on $H_{0}$ changes slowly with magnitude to even brighter magnitude limits, even down to $r=11$ for LSST. We stop plotting results below this because there are only 700 galaxies being used and some of the redshift bins for individual surveys start becoming empty because
of Poisson fluctuations.  The NGRST HLS result cuts off at $r=13$ and below, for the same reason  (there are a similar number of galaxies per survey at this limiting magnitude).  The shaded regions around the lines show the
statistical $1 \sigma$ error on the $\sigma_{H_{0}}$ values, computed from the scatter between results for the
different mock surveys. Adding galaxies fainter than $r=18$ to the sample does not noticeably increase the precision of the
$H_{0}$ measurement, which remains within the 1 $\sigma$ error on $\sigma_{H_{0}}$  for both LSST and NGRST HLS.

As for Section \ref{resultsastroerr}, we have also computed a curves for analyses where the resolution elements of 
galaxies are not used independently. These are shown as a blue lines in each of the LSST and NGRST HLS panels. We can see that even extending the number of galaxies used by a large factor with a magnitude limit of $r=20$ (which results in 1.8 million galaxies on average per LSST mock) does
not decrease the error on $H_{0}$ to below 10\% for LSST, indicating that techniques to make use of separate resolved elements will be essential in this case for a competitive measurement. For NGRST, the situation is better, and although $\sigma_{H_{0}}=6.8\%$ for the fiducial $r=18$ magnitude limit, it
does decrease to $\sigma_{H_{0}}=1.5\%$ for $r=20$. Of course in this case it will still be necessary to
make one measurement for each galaxy, which are extended objects and so this will not be straightforward.

\subsection{Impact of independent image elements and velocity reconstruction}
\label{elems}
\begin{figure}
 \includegraphics[width=1.0\columnwidth]{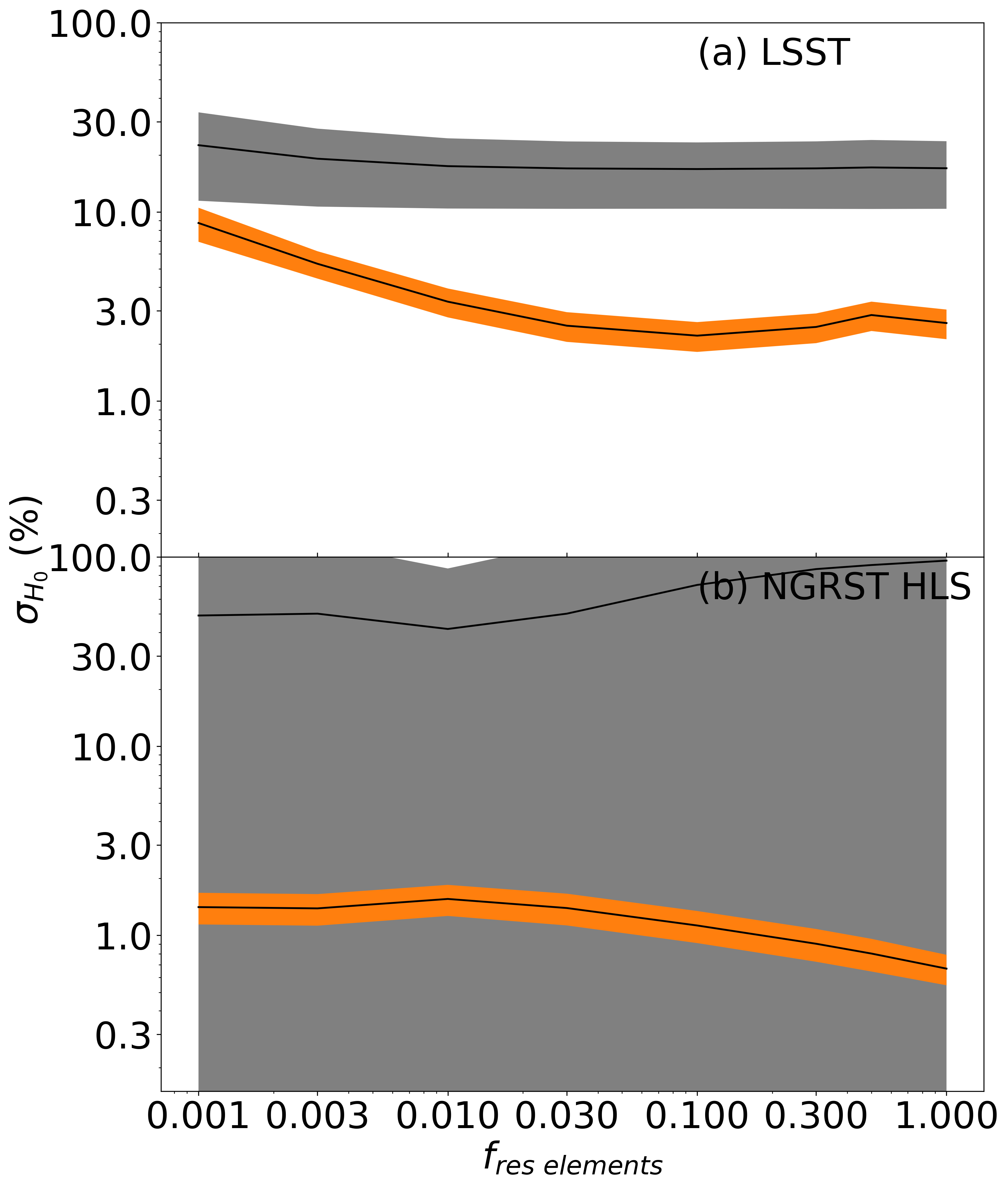}
 \caption{
 The {\it rms} percentage error on $H_{0}$ measurements as a function of the parameter $f_{\rm res \, elem}$ in Equation \ref{astroerr2}, which governs the fraction of resolution elements in each galaxy which are used to make independent astrometric measurements.
  The results shown in orange are for the case where the rest of the fiducial analysis is kept the same (including application of a velocity flow model to 
  correct peculiar velocities). The results in gray are for an analysis with no peculiar velocity corrections.
  The envelope around each line illustrates  the standard deviation of the  $\sigma_{H_{0}}$ values for
the mock surveys. The top panel, (a) is for mock LSSTs, and the bottom panel, 
(b) is for mock NGRST HLSs.
}
 \label{elementfac}
\end{figure}

How well one can use the individual resolved elements of galaxies to improve the astrometric measurement precision
is an open question. Future work should test this with simulations and tests with observational data. In the meantime,
we have varied the parameter
$f_{\rm res \, elements}$ described in Section \ref{fida} to quantify the how the fraction of resolved elements
that are used independently affects the overall measurement error on $H_{0}$. In Figure \ref{elementfac} we 
show the results of varying $f_{\rm res} {\rm elements}$ from 0.001 to 1.0. With our fiduc ial value, 
$f_{\rm res \, elements}=0.5$ we have the values of $\sigma_{H_{0}}=2.8$\% (LSST)
and  $\sigma_{H_{0}}=0.8$\% (NGRST HLS) quoted above. This decreases to
 $\sigma_{H_{0}}=2.6$\% (LSST) and $\sigma_{H_{0}}=0.7$\% (NGRST HLS) when $f_{\rm res \, elements}=1.0$. Although  $\sigma_{H_{0}}$ increases 
 as $f_{\rm res \, elements}$ decreases, even when $f_{\rm res} {\rm elements}=0.001$, $\sigma_{H_{0}}$ is
  significantly better than for the case when each galaxy is only used to make a single
 measurement (see Figure \ref{rmaglim}). We find that bright LSST galaxies with $r\sim 15$
 at redshifts $z\sim0.03$ (about halfway to the far mock boundary) typically contain
$\sim 3000$ resolved elements. If this can be exploited, it is this which will give the astrometry measurements their power.

In  the same figure, \ref{elementfac} we also show the effect of not including any peculiar velocity flow modeling, as grey curves. We can see that the peculiar velocities dominate the error on $H_{0}$, and even when using all
elements of resolved galaxies, $f_{\rm res} {\rm elements}=1.0$, the value of $\sigma_{H_{0}}$ stays at about 20\% for the LSST. For NGRST HLS, because the mock volumes are an order of magnitude smaller than for LSST, the effect of not correcting for peculiar velocities is much worse, with $\sigma_{H_{0}}$ being around 100\%.
All components of our fiducial modeling will therefore have to be applied and work well if the technique is to
yield a competitive measurement (or even a  detection in the case of NGRST).

\section{Summary and discussion}

\label{summary}

\subsection{Summary}
We have investigated the possible use of either the Vera Rubin Observatory's galaxy survey (the
LSST), or the Nancy Grace Roman Space Telescope HLS to measure the parallax shift of galaxies below $z=0.06$ due to the Earth's motion with respect
to the CMB frame, and what the resulting error bars on a Hubble constant constraint from such a
measurement might be. To generate our predictions 
we have used N-body simulations to make mock catalogues and included
the expected astrometric errors and the effects of peculiar velocities. We have also made various
assumptions about how the resolved elements of galaxies could be used to reduce the astrometric
errors and about the correction of peculiar velocities with a flow model.  Our conclusions are
as follows:
\begin{enumerate}
    \item With our fiducial assumptions and analysis techniques, we find
a $2.8\%$ error on $H_{0}$ from our analysis of LSST mock surveys and
$0.8 \%$ error for NGRST HLS.
\item  In order to achieve competitive results, the individual 
resolved elements of nearby large galaxies will have to be used  for
astrometry in a way that leads to a $\sim \sqrt{N}$ reduction in the
errors where $N$ is the number of elements per galaxy. This will
require development of new techniques. Some simple preliminary tests using 
observational galaxy data in Appendix A are relatively promising, but their relevance to the eventual measurements is unclear.
Without this averaging over
resolved elements, the error on $H_0$ is $\sim 10-30\%$.
\item  Proper motions from coherent peculiar velocities are a strong
contaminant to the measurement, to such
an extent that a flow field model will be essential to correct them. Without
this the error on $H_0$ is $\sim 20\%$ for LSST and $50-100\%$ for NGRST HLS
\item It will be necessary to obtain spectroscopic redshifts  for a significant
subsample of galaxies, primarily to enable the use of flow model corrections. Our
fiducial apparent magnitude limit for spectroscopic redshifts is $r=18$, which would
mean measurments for about $400,000$ galaxies for LSST or $40,000$ for NGRST HLS. To ensure an $H_0$ error of $<3\%$
we require $r<14$ for both LSST and NGRST.
\end{enumerate}

\subsection{Discussion}
Although our conclusions are promising, it is obvious that in order
to succeed, and make the first truly geometrical measurement
of $H_{0}$, it will be necessary to carry out a very large program, and
overcome many difficulties.  We have been able only to sketch out
in general terms what will need to be achieved in various areas (e.g.,
peculiar velocity reconstruction), and future work could uncover 
further complications, or find that additional observational or systematic
errors appear which make the measurement much more difficult. 

Realistically, one can expect the precision of other Hubble constant
measurement techniques (such as cepheids) to increase. For example, some 
current error bars described in \cite{riess2019b} are
of order 1\%. This implies that even if an optimistic
parallax measurement case (that we have outlined in this paper) plays out,
then it is not likely to be close in statistical precision to other methods. 
Because the parallax method depends on so many variables (for example even the Gaia satellite developed for astrometry has been affected by unanticipated field distortions), it seems most likely that the culmination of
parallax measurement in the forseeable future will be a detection of
the effect rather than a precise determination. Of course this in itself would be ground breaking and worth pursuing.

The LSST case appears to be the most uncertain, as optical ground based astrometry is intrinsically much more difficult than that carried out in
space. Even with satellites such as Gaia, it has been difficult to achieve the planned 
measurement accuracy. For example, in the second year data release \citep{gaiadr2}, residual
errors of $\sim$milliarcsec $yr^{-1}$, coherent over angular scales of $\sim0.5-1$ deg. have complicated analyses \citep[e.g.,][]{vas19}.  
The science requirements for the LSST specify a proper motion error of 0.2 milliarcsec yr$^{-1}$,
for bright point sources. An example of ground based astrometry is the measurement of
the proper motion of the Fornax local group galaxy by \cite{mendez11}, with 
the two orthogonal components measured being $0.62\pm0.16$ milliarcsec yr$^{-1}$ and $0.53\pm0.15$
milliarcsec yr$^{-1}$. We have assumed here that over the lifetime of the LSST the science
requirements will be achievable.

The precision achieved by \cite{mendez11} involved combining measurements for individual stars in the
Fornax galaxy.  Most of the galaxies which would be used in our proposed study for both 
Rubin Observatory and NGRST would not have individually resolved stars, but as we have seen it will be necessary to improve the astrometry measurements beyond
one measurement per galaxy. The type of techniques that could be developed to achieve this could
involve cross-correlation of shifted whole galaxy images, or perhaps identification of 
individual bright maxima in resolved surface brightness maps. It is certain that this will require
significant  effort and new ideas to bring about. The systematic
errors in astrometric measurement detailed by e.g., \cite{sanderson17} for
NGRST in the context of point sources (such as pixel placement error) are likely to become even more difficult to deal with for extended objects.

These new techniques for extended object astrometry will need to be tested on both real and simulated data.
In this paper, we have also assumed that the precision of point source astrometry can be applied to the individual resolved elements of
extended objects. This will not be true in detail, and may degrade
the precision that can be achieved.

We have seen that one of the most crucial aspects of the analysis is the use of a flow field
model to correct for the proper motions of galaxies. This is another area which will need to
be tested in depth. In our present work, we have merely constructed the flow field model
from the smoothed simulation velocities, and added a random velocity dispersion term. In the
future, techniques to predict the velocities from the galaxy distribution (such as e.g.,
\citealt{keselman17}), should be used and tested with simulations. 
Peculiar velocity surveys derived from standard candle based distance indicators (e.g., \citealt{graziani19}) could also be useful.
If the residual velocity errors are
randomly distributed, then we find that they would have no detectable impact. For example, the extra
velocity dispersion term we add to the flow model to mimic velocity reconstruction errors is
$100 \kms$ in the fiducial analysis. If we increase this to $1000 \kms$, we find no change in the
error on the Hubble constant. On  the other hand, more realistic velocity reconstruction errors are likely to have some spatial coherence, and how this will affect the error budget should be evaluated
with realistic simulations. Also, as we have mentioned in Section \ref{flowmodel}, statistical
measurement of proper motions from the astrometry should allow the flow model itself to be tested.

Our mock surveys have been drawn from a large simulation of a CDM universe, from 27 sites which are effectively chosen  randomly with respect to the large scale
structures present. In the case of an observational measurement, the neighbourhood of the Milky Way is a very particular environment, and one that we have not
attempted to model. This will affect our estimates of the error bars on the $H_{0}$ measurement, which we have seen from Figure \ref{errhist} can vary by a factor of 3 among the different mocks. For example, the local galaxy velocity field may be colder or
hotter than average, or the Local Group may be a denser environment than that surrounding observers in the majority of our mocks. As a further simplification, we have decoupled the actual cosmic velocity field at the observer's position in our mocks from the value used to compute the cosmic secular parallax in Equation \ref{pieq} (where we use the Local Group's observed value). In a truly realistic simulation, these should be consistent.  To address these issues when
making theoretical predictions, one could use
constrained realization simulations (\citealt{hoffman91,clues16}), where the observer position is forced to have some of the same characteristics as the local group,
or else mock observers with the right properties could be picked from 
larger simulations. In the end, when dealing with observations, one will be forced to deal with a single example, the Local Group
itself, and flow field models specifically generated in that context will be used
(e.g., \cite{paine20}).

More accurate forecasts will also entail the use of better models for the sizes of galaxies,
and their luminosities. We have used an extremely crude conversion of dark matter subhalo mass
to galaxy light using a constant mass to light ratio. In the future, techniques such as
subhalo abundance matching (e.g., \citealt{conroy06}) could be useful to ensure that the mock surveys more
closely match the luminosity distribution of galaxies in the real Universe. The sizes of galaxies,
used to estimate the number of resolved elements per galaxy, were also gauged very roughly and
there is definite room for improvement there, such as including the effect of the galaxy
luminosity profile.

Apart from observational and theoretical systematic errors, systematic and statistical
errors from astrophysical effects will also arise. For example stellar or AGN variability 
could cause galaxy centroid shifts, or at least angular  shifts in  the positions of bright
resolved galaxy elements. Other sources of large coherent proper motions such as the secular acceleration of the solar system with respect to the Milky Way (e.g., \citealt{kopeikin06}) 
will need to be accounted for.
Gravitational waves can cause apparent proper motions, and astrometric measurements will be sensitive to those also \citep{darling2018,paine20}. 

Although many of the issues discussed above could result in degradation of the $H_{0}$ errors,
one aspect which could improve constraints is increasing the outer redshift boundary of
the galaxy survey used. In our present work, we were limited  by the size of the simulation
volume and the necessity to create many mock surveys to an outer boundary of $z=0.058$ ($175 \hmpc$
from the observer). We have seen (e.g.,  Figure \ref{spectroz}) that the cosmic secular parallax
is in principle detectable at that redshift or beyond. For example DC examined
what might be possible with quasar parallax (albeit with different, more futuristic instruments), and considered data from quasars at redshifts both up to and beyond $z=2$ (which could be used to constrain dark energy). In the case of the Rubin Observatory and NGRST, if the effect could be measured at  $z=0.058$, it seems likely
that it could be extended, and at higher redshifts  photometric redshifts 
(with their smaller fractional errors)
might also play a larger role.

Of the two telescopes we have considered, Rubin Observatory and NGRST, it is clear that NGRST has many advantages. Apart from avoiding the
difficulty of ground based astrometry,  NGRST will operate in the infrared, a wavelength regime where the most precise
relative astrometry has so far been achieved from the ground (with adaptive optics, 150 $\mu$ arcsec:  \citealt{ghez08}). The estimated proper motion errors from 
the HLS for bright sources are a factor of 8 smaller than for LSST. We have seen
that this leads to the fractional error on $H_{0}$ being a factor of 3.5 smaller.
 We have also seen in this paper that for nearby ($z<0.06$ )
galaxies, proper motions from the cosmic peculiar velocity field are the greatest
limiting precision to $H_{0}$ measurement from galaxy astrometry. Use of  NGRST 
rather than Rubin Observatory is unlikely to help directly with this aspect, although 
the smaller astrometric errors  may allow galaxies at greater distances
to be used. Nevertheless, it would be very useful to attempt the project with both instruments. The LSST covers a much larger sky area, and so covers directions where the
secular parallax will be non existent, and areas where it will be maximal. Many of the systematic effects dealt with will be different between the two instruments, and it will be possible to
compare the measurements in order to verify the overall result. In addition, the achievable statistical precision on $H_{0}$, of below $3 \%$ even for the LSST is enough to make the end result interesting in itself.

Cosmic secular parallax, like the cosmic redshift drift \citep{sandage62,loeb1998} is a
"real-time cosmology" effect. A measurement is also currently out of reach, but as we have 
shown,  instruments that may be capable of making one, either the Rubin Observatory, or NGRST
are both currently nearing completion. The endeavour will require
advances in the astrometry of extended objects, velocity field modelling, and many other areas,
but the payoff will be great.  Being
able to directly detect the parallax shifts of galaxies at cosmic distances would fulfill a
fundamental goal of cosmology, direct determination of the scale of the Universe.

\subsection*{Data availability}
The data underlying this article will be shared on reasonable request to the corresponding author. The datasets were derived from sources in the public domain: {\tt https://hpc.imit.chiba-u.jp/$\sim$ishiymtm/db.html}, {\tt https://archive.stsci.edu/ } and {\tt https://nsatlas.org}.

\subsection*{Acknowledgments}
RACC thanks Michael Pierce, Sergey Koposov, Karl Glazebrook and Douglas Clowe
for useful discussions, and thanks Stuart Wyithe and the University of Melbourne for
their hospitality. 
RACC acknowledges support from  NASA  ATP 80NSSC18K101,  NASA ATP
NNX17AK56G, NSF  AST-1909193, and a Lyle fellowship from the University of Melbourne.
RACC thanks Tomoaki Ishiyama for making simulation data publicly available.

\bibliographystyle{mnras}
\bibliography{main}

\begin{thebibliography}{}
\makeatletter
\relax
\def\mn@urlcharsother{\let\do\@makeother \do\$\do\&\do\#\do\^\do\_\do\%\do\~}
\def\mn@doi{\begingroup\mn@urlcharsother \@ifnextchar [ {\mn@doi@}
  {\mn@doi@[]}}
\def\mn@doi@[#1]#2{\def\@tempa{#1}\ifx\@tempa\@empty \href
  {http://dx.doi.org/#2} {doi:#2}\else \href {http://dx.doi.org/#2} {#1}\fi
  \endgroup}
\def\mn@eprint#1#2{\mn@eprint@#1:#2::\@nil}
\def\mn@eprint@arXiv#1{\href {http://arxiv.org/abs/#1} {{\tt arXiv:#1}}}
\def\mn@eprint@dblp#1{\href {http://dblp.uni-trier.de/rec/bibtex/#1.xml}
  {dblp:#1}}
\def\mn@eprint@#1:#2:#3:#4\@nil{\def\@tempa {#1}\def\@tempb {#2}\def\@tempc
  {#3}\ifx \@tempc \@empty \let \@tempc \@tempb \let \@tempb \@tempa \fi \ifx
  \@tempb \@empty \def\@tempb {arXiv}\fi \@ifundefined
  {mn@eprint@\@tempb}{\@tempb:\@tempc}{\expandafter \expandafter \csname
  mn@eprint@\@tempb\endcsname \expandafter{\@tempc}}}

\bibitem[\protect\citeauthoryear{{Abbott} et~al.,}{{Abbott}
  et~al.}{2017}]{abbott17}
{Abbott} B.~P.,  et~al., 2017, \mn@doi [\prl] {10.1103/PhysRevLett.119.161101},
  \href {https://ui.adsabs.harvard.edu/abs/2017PhRvL.119p1101A} {119, 161101}

\bibitem[\protect\citeauthoryear{{Abell} et~al.,}{{Abell}
  et~al.}{2009}]{sciencebook}
{Abell} P.~A.,  et~al., 2009, arXiv e-prints arXiv:0912.0201, \href
  {https://ui.adsabs.harvard.edu/abs/2009arXiv0912.0201L} {}

\bibitem[\protect\citeauthoryear{{Abitbol}, {Hill}  \& {Chluba}}{{Abitbol}
  et~al.}{2020}]{abitbol}
{Abitbol} M.~H.,  {Hill} J.~C.,   {Chluba} J.,  2020, \mn@doi [\apj]
  {10.3847/1538-4357/ab7b70}, \href
  {https://ui.adsabs.harvard.edu/abs/2020ApJ...893...18A} {893, 18}

\bibitem[\protect\citeauthoryear{{Ade} et~al.,}{{Ade}
  et~al.}{2014}]{planck2014}
{Ade} P.~A.~R.,  et~al., 2014, \mn@doi [\aap] {10.1051/0004-6361/201321591},
  \href {https://ui.adsabs.harvard.edu/abs/2014A&A...571A..16P} {571, A16}

\bibitem[\protect\citeauthoryear{{Akrami} et~al.,}{{Akrami}
  et~al.}{2020}]{planck20}
{Akrami} Y.,  et~al., 2020, arXiv e-prints arXiv:2003.12646, \href
  {https://ui.adsabs.harvard.edu/abs/2020arXiv200312646P} {}

\bibitem[\protect\citeauthoryear{{Bahcall} \& {Kulier}}{{Bahcall} \&
  {Kulier}}{2014}]{bahcall14}
{Bahcall} N.~A.,  {Kulier} A.,  2014, \mn@doi [\mnras] {10.1093/mnras/stu107},
  \href {https://ui.adsabs.harvard.edu/abs/2014MNRAS.439.2505B} {439, 2505}

\bibitem[\protect\citeauthoryear{{Bailer-Jones}, {Rybizki}, {Fouesneau},
  {Mantelet}  \& {Andrae}}{{Bailer-Jones} et~al.}{2018}]{bailer2018}
{Bailer-Jones} C.~A.~L.,  {Rybizki} J.,  {Fouesneau} M.,  {Mantelet} G.,
  {Andrae} R.,  2018, \mn@doi [\aj] {10.3847/1538-3881/aacb21}, \href
  {https://ui.adsabs.harvard.edu/abs/2018AJ....156...58B} {156, 58}

\bibitem[\protect\citeauthoryear{{Behroozi}, {Wechsler}  \& {Wu}}{{Behroozi}
  et~al.}{2013}]{rockstar}
{Behroozi} P.~S.,  {Wechsler} R.~H.,   {Wu} H.-Y.,  2013, \mn@doi [\apj]
  {10.1088/0004-637X/762/2/109}, \href
  {https://ui.adsabs.harvard.edu/abs/2013ApJ...762..109B} {762, 109}

\bibitem[\protect\citeauthoryear{{Beutler} et~al.,}{{Beutler}
  et~al.}{2011}]{beutler11}
{Beutler} F.,  et~al., 2011, \mn@doi [\mnras]
  {10.1111/j.1365-2966.2011.19250.x}, \href
  {https://ui.adsabs.harvard.edu/abs/2011MNRAS.416.3017B} {416, 3017}

\bibitem[\protect\citeauthoryear{{Binney} \& {Merrifield}}{{Binney} \&
  {Merrifield}}{1998}]{binney98}
{Binney} J.,  {Merrifield} M.,  1998, {Galactic Astronomy}.
Princeton

\bibitem[\protect\citeauthoryear{{Blanton}, {Eisenstein}, {Hogg}, {Schlegel}
  \& {Brinkmann}}{{Blanton} et~al.}{2005}]{blanton05}
{Blanton} M.~R.,  {Eisenstein} D.,  {Hogg} D.~W.,  {Schlegel} D.~J.,
  {Brinkmann} J.,  2005, \mn@doi [\apj] {10.1086/422897}, \href
  {https://ui.adsabs.harvard.edu/abs/2005ApJ...629..143B} {629, 143}

\bibitem[\protect\citeauthoryear{{Boehm} et~al.,}{{Boehm} et~al.}{2017}]{theia}
{Boehm} C.,  et~al., 2017, arXiv e-prints arXiv:1707.01348, \href
  {https://ui.adsabs.harvard.edu/abs/2017arXiv170701348T} {}

\bibitem[\protect\citeauthoryear{{Bonifacio} et~al.,}{{Bonifacio}
  et~al.}{2016}]{weave16}
{Bonifacio} P.,  et~al., 2016, in {Reyl{\'e}} C.,  {Richard} J.,
  {Cambr{\'e}sy} L.,  {Deleuil} M.,  {P{\'e}contal} E.,  {Tresse} L.,
  {Vauglin} I.,  eds, SF2A-2016: Proceedings of the Annual meeting of the
  French Society of Astronomy and Astrophysics. pp 267--270

\bibitem[\protect\citeauthoryear{{Branchini} et~al.,}{{Branchini}
  et~al.}{1999}]{branchini99}
{Branchini} E.,  et~al., 1999, \mn@doi [\mnras]
  {10.1046/j.1365-8711.1999.02514.x}, \href
  {https://ui.adsabs.harvard.edu/abs/1999MNRAS.308....1B} {308, 1}

\bibitem[\protect\citeauthoryear{{Brown} et~al.,}{{Brown}
  et~al.}{2018}]{gaiadr2}
{Brown} A.~G.~A.,  et~al., 2018, \mn@doi [\aap] {10.1051/0004-6361/201833051},
  \href {https://ui.adsabs.harvard.edu/abs/2018A&A...616A...1G} {616, A1}

\bibitem[\protect\citeauthoryear{{Carlesi} et~al.,}{{Carlesi}
  et~al.}{2016}]{clues16}
{Carlesi} E.,  et~al., 2016, \mn@doi [\mnras] {10.1093/mnras/stw357}, \href
  {https://ui.adsabs.harvard.edu/abs/2016MNRAS.458..900C} {458, 900}

\bibitem[\protect\citeauthoryear{{Chen} et~al.,}{{Chen}
  et~al.}{2019}]{chen2019}
{Chen} G. C.~F.,  et~al., 2019, \mn@doi [\mnras] {10.1093/mnras/stz2547}, \href
  {https://ui.adsabs.harvard.edu/abs/2019MNRAS.490.1743C} {490, 1743}

\bibitem[\protect\citeauthoryear{{Chisari} et~al.,}{{Chisari}
  et~al.}{2019}]{chisari2019}
{Chisari} N.~E.,  et~al., 2019, \mn@doi [\apjs] {10.3847/1538-4365/ab1658},
  \href {https://ui.adsabs.harvard.edu/abs/2019ApJS..242....2C} {242, 2}

\bibitem[\protect\citeauthoryear{{Colombi}, {Chodorowski}  \&
  {Teyssier}}{{Colombi} et~al.}{2007}]{colombi07}
{Colombi} S.,  {Chodorowski} M.~J.,   {Teyssier} R.,  2007, \mn@doi [\mnras]
  {10.1111/j.1365-2966.2006.11330.x}, \href
  {https://ui.adsabs.harvard.edu/abs/2007MNRAS.375..348C} {375, 348}

\bibitem[\protect\citeauthoryear{{Conroy}, {Wechsler}  \& {Kravtsov}}{{Conroy}
  et~al.}{2006}]{conroy06}
{Conroy} C.,  {Wechsler} R.~H.,   {Kravtsov} A.~V.,  2006, \mn@doi [\apj]
  {10.1086/503602}, \href
  {https://ui.adsabs.harvard.edu/abs/2006ApJ...647..201C} {647, 201}

\bibitem[\protect\citeauthoryear{{Croft} \& {Dailey}}{{Croft} \&
  {Dailey}}{2011}]{croft11}
{Croft} R. A.~C.,  {Dailey} M.,  2011, arXiv e-prints arXiv:1112.3108, \href
  {https://ui.adsabs.harvard.edu/abs/2011arXiv1112.3108C} {}

\bibitem[\protect\citeauthoryear{{Croft} \& {Gaztanaga}}{{Croft} \&
  {Gaztanaga}}{1997}]{croft97}
{Croft} R. A.~C.,  {Gaztanaga} E.,  1997, \mn@doi [\mnras]
  {10.1093/mnras/285.4.793}, \href
  {https://ui.adsabs.harvard.edu/abs/1997MNRAS.285..793C} {285, 793}

\bibitem[\protect\citeauthoryear{{Crossland}, {Stenetorp}, {Riedel}, {Kawata},
  {Kitching}  \& {Croft}}{{Crossland} et~al.}{2020}]{crossland20}
{Crossland} T.,  {Stenetorp} P.,  {Riedel} S.,  {Kawata} D.,  {Kitching} T.~D.,
    {Croft} R. A.~C.,  2020, \mn@doi [\mnras] {10.1093/mnras/stz3400}, \href
  {https://ui.adsabs.harvard.edu/abs/2020MNRAS.492.3217C} {492, 3217}

\bibitem[\protect\citeauthoryear{{Cuceu}, {Farr}, {Lemos}  \&
  {Font-Ribera}}{{Cuceu} et~al.}{2019}]{cuceu19}
{Cuceu} A.,  {Farr} J.,  {Lemos} P.,   {Font-Ribera} A.,  2019, \mn@doi [\jcap]
  {10.1088/1475-7516/2019/10/044}, \href
  {https://ui.adsabs.harvard.edu/abs/2019JCAP...10..044C} {2019, 044}

\bibitem[\protect\citeauthoryear{{Darling}}{{Darling}}{2012}]{darling2012}
{Darling} J.,  2012, \mn@doi [\apjl] {10.1088/2041-8205/761/2/L26}, \href
  {https://ui.adsabs.harvard.edu/abs/2012ApJ...761L..26D} {761, L26}

\bibitem[\protect\citeauthoryear{{Darling} \& {Truebenbach}}{{Darling} \&
  {Truebenbach}}{2018}]{darling2018}
{Darling} J.,  {Truebenbach} A.~E.,  2018, \mn@doi [\apj]
  {10.3847/1538-4357/aad3d0}, \href
  {https://ui.adsabs.harvard.edu/abs/2018ApJ...864...37D} {864, 37}

\bibitem[\protect\citeauthoryear{{Darling}, {Truebenbach}  \&
  {Paine}}{{Darling} et~al.}{2018}]{darling18}
{Darling} J.,  {Truebenbach} A.~E.,   {Paine} J.,  2018, \mn@doi [\apj]
  {10.3847/1538-4357/aac772}, \href
  {https://ui.adsabs.harvard.edu/abs/2018ApJ...861..113D} {861, 113}

\bibitem[\protect\citeauthoryear{{Davis} et~al.,}{{Davis}
  et~al.}{2011}]{davis11}
{Davis} T.~M.,  et~al., 2011, \mn@doi [\apj] {10.1088/0004-637X/741/1/67},
  \href {https://ui.adsabs.harvard.edu/abs/2011ApJ...741...67D} {741, 67}

\bibitem[\protect\citeauthoryear{{Dhawan}, {Jha}  \& {Leibundgut}}{{Dhawan}
  et~al.}{2018}]{dhawan18}
{Dhawan} S.,  {Jha} S.~W.,   {Leibundgut} B.,  2018, \mn@doi [\aap]
  {10.1051/0004-6361/201731501}, \href
  {https://ui.adsabs.harvard.edu/abs/2018A&A...609A..72D} {609, A72}

\bibitem[\protect\citeauthoryear{{Ding} \& {Croft}}{{Ding} \&
  {Croft}}{2009}]{ding2009}
{Ding} F.,  {Croft} R. A.~C.,  2009, \mn@doi [\mnras]
  {10.1111/j.1365-2966.2009.15111.x}, \href
  {https://ui.adsabs.harvard.edu/abs/2009MNRAS.397.1739D} {397, 1739}

\bibitem[\protect\citeauthoryear{{Freedman} et~al.,}{{Freedman}
  et~al.}{2001}]{freedman01}
{Freedman} W.~L.,  et~al., 2001, \mn@doi [\apj] {10.1086/320638}, \href
  {https://ui.adsabs.harvard.edu/abs/2001ApJ...553...47F} {553, 47}

\bibitem[\protect\citeauthoryear{{Freedman} et~al.,}{{Freedman}
  et~al.}{2019}]{freedman2019}
{Freedman} W.~L.,  et~al., 2019, \mn@doi [\apj] {10.3847/1538-4357/ab2f73},
  \href {https://ui.adsabs.harvard.edu/abs/2019ApJ...882...34F} {882, 34}

\bibitem[\protect\citeauthoryear{{Ghez} et~al.,}{{Ghez} et~al.}{2008}]{ghez08}
{Ghez} A.~M.,  et~al., 2008, \mn@doi [\apj] {10.1086/592738}, \href
  {https://ui.adsabs.harvard.edu/abs/2008ApJ...689.1044G} {689, 1044}

\bibitem[\protect\citeauthoryear{{Gorski}, {Davis}, {Strauss}, {White}  \&
  {Yahil}}{{Gorski} et~al.}{1989}]{gorski89}
{Gorski} K.~M.,  {Davis} M.,  {Strauss} M.~A.,  {White} S. D.~M.,   {Yahil} A.,
   1989, \mn@doi [\apj] {10.1086/167771}, \href
  {https://ui.adsabs.harvard.edu/abs/1989ApJ...344....1G} {344, 1}

\bibitem[\protect\citeauthoryear{{Graham}}{{Graham}}{2019}]{graham2019}
{Graham} M.,  2019, in The Extragalactic Explosive Universe: the New Era of
  Transient Surveys and Data-Driven Discovery. p.~23,
  \mn@doi{10.5281/zenodo.3478038}

\bibitem[\protect\citeauthoryear{{Graham}, {Connolly}, {Ivezi{\'c}}, {Schmidt},
  {Jones}, {Juri{\'c}}, {Daniel}  \& {Yoachim}}{{Graham}
  et~al.}{2018}]{graham18}
{Graham} M.~L.,  {Connolly} A.~J.,  {Ivezi{\'c}} {\v{Z}}.,  {Schmidt} S.~J.,
  {Jones} R.~L.,  {Juri{\'c}} M.,  {Daniel} S.~F.,   {Yoachim} P.,  2018,
  \mn@doi [\aj] {10.3847/1538-3881/aa99d4}, \href
  {https://ui.adsabs.harvard.edu/abs/2018AJ....155....1G} {155, 1}

\bibitem[\protect\citeauthoryear{{Gramann}, {Cen}  \& {Gott}}{{Gramann}
  et~al.}{1994}]{gramann94}
{Gramann} M.,  {Cen} R.,   {Gott} J.~Richard I.,  1994, \mn@doi [\apj]
  {10.1086/173994}, \href
  {https://ui.adsabs.harvard.edu/abs/1994ApJ...425..382G} {425, 382}

\bibitem[\protect\citeauthoryear{{Graziani}, {Courtois}, {Lavaux}, {Hoffman},
  {Tully}, {Copin}  \& {Pomar{\`e}de}}{{Graziani} et~al.}{2019}]{graziani19}
{Graziani} R.,  {Courtois} H.~M.,  {Lavaux} G.,  {Hoffman} Y.,  {Tully} R.~B.,
  {Copin} Y.,   {Pomar{\`e}de} D.,  2019, \mn@doi [\mnras]
  {10.1093/mnras/stz078}, \href
  {https://ui.adsabs.harvard.edu/abs/2019MNRAS.488.5438G} {488, 5438}

\bibitem[\protect\citeauthoryear{Guizar, Thurman  \& Fienup}{Guizar
  et~al.}{2008}]{guizar08}
Guizar M.,  Thurman S.,   Fienup J.,  2008, Optics Letters, 33, 156

\bibitem[\protect\citeauthoryear{{Hall}}{{Hall}}{2019}]{hall19}
{Hall} A.,  2019, \mn@doi [\mnras] {10.1093/mnras/stz648}, \href
  {https://ui.adsabs.harvard.edu/abs/2019MNRAS.486..145H} {486, 145}

\bibitem[\protect\citeauthoryear{{Hinshaw} et~al.,}{{Hinshaw}
  et~al.}{2009}]{hinshaw09}
{Hinshaw} G.,  et~al., 2009, \mn@doi [\apjs] {10.1088/0067-0049/180/2/225},
  \href {https://ui.adsabs.harvard.edu/abs/2009ApJS..180..225H} {180, 225}

\bibitem[\protect\citeauthoryear{{Hoffman} \& {Ribak}}{{Hoffman} \&
  {Ribak}}{1991}]{hoffman91}
{Hoffman} Y.,  {Ribak} E.,  1991, \mn@doi [\apjl] {10.1086/186160}, \href
  {https://ui.adsabs.harvard.edu/abs/1991ApJ...380L...5H} {380, L5}

\bibitem[\protect\citeauthoryear{{Hogg}}{{Hogg}}{1999}]{hogg99}
{Hogg} D.~W.,  1999, arXiv e-prints astro-ph/9905116, \href
  {https://ui.adsabs.harvard.edu/abs/1999astro.ph..5116H} {}

\bibitem[\protect\citeauthoryear{{Holz} \& {Hughes}}{{Holz} \&
  {Hughes}}{2005}]{holz2005}
{Holz} D.~E.,  {Hughes} S.~A.,  2005, \mn@doi [\apj] {10.1086/431341}, \href
  {https://ui.adsabs.harvard.edu/abs/2005ApJ...629...15H} {629, 15}

\bibitem[\protect\citeauthoryear{{Howlett} \& {Davis}}{{Howlett} \&
  {Davis}}{2020}]{howlett20}
{Howlett} C.,  {Davis} T.~M.,  2020, \mn@doi [\mnras] {10.1093/mnras/staa049},
  \href {https://ui.adsabs.harvard.edu/abs/2020MNRAS.492.3803H} {492, 3803}

\bibitem[\protect\citeauthoryear{{Hrazd{\'\i}ra}, {Druckm{\"u}ller}  \&
  {Habbal}}{{Hrazd{\'\i}ra} et~al.}{2020}]{hrazd20}
{Hrazd{\'\i}ra} Z.,  {Druckm{\"u}ller} M.,   {Habbal} S.,  2020, \mn@doi
  [\apjs] {10.3847/1538-4365/ab63d7}, \href
  {https://ui.adsabs.harvard.edu/abs/2020ApJS..247....8H} {247, 8}

\bibitem[\protect\citeauthoryear{{Hubble}}{{Hubble}}{1925}]{hubble1925}
{Hubble} E.~P.,  1925, \mn@doi [\apj] {10.1086/142943}, \href
  {https://ui.adsabs.harvard.edu/abs/1925ApJ....62..409H} {62, 409}

\bibitem[\protect\citeauthoryear{{Huchra}}{{Huchra}}{1992}]{huchra92}
{Huchra} J.~P.,  1992, \mn@doi [Science] {10.1126/science.256.5055.321}, \href
  {https://ui.adsabs.harvard.edu/abs/1992Sci...256..321H} {256, 321}

\bibitem[\protect\citeauthoryear{{Ishiyama}, {Enoki}, {Kobayashi}, {Makiya},
  {Nagashima}  \& {Oogi}}{{Ishiyama} et~al.}{2015}]{ishiyama15}
{Ishiyama} T.,  {Enoki} M.,  {Kobayashi} M. A.~R.,  {Makiya} R.,  {Nagashima}
  M.,   {Oogi} T.,  2015, \mn@doi [\pasj] {10.1093/pasj/psv021}, \href
  {https://ui.adsabs.harvard.edu/abs/2015PASJ...67...61I} {67, 61}

\bibitem[\protect\citeauthoryear{{Ivesi\'c} et~al.}{{Ivesi\'c}
  et~al.}{2018}]{ivesic18}
{Ivesi\'c} Z.,  et~al., 2018, LSST document LPM-17, pp LPM--17

\bibitem[\protect\citeauthoryear{{Ivezi{\'c}}, {Beers}  \&
  {Juri{\'c}}}{{Ivezi{\'c}} et~al.}{2012}]{ivesic12}
{Ivezi{\'c}} {\v{Z}}.,  {Beers} T.~C.,   {Juri{\'c}} M.,  2012, \mn@doi [\araa]
  {10.1146/annurev-astro-081811-125504}, \href
  {https://ui.adsabs.harvard.edu/abs/2012ARA&A..50..251I} {50, 251}

\bibitem[\protect\citeauthoryear{{Kardashev}, {Parijskij}  \&
  {Umarbaeva}}{{Kardashev} et~al.}{1973}]{kardashev73}
{Kardashev} N.~S.,  {Parijskij} Y.~N.,   {Umarbaeva} N.~D.,  1973,
  Astrofizicheskie Issledovaniia Izvestiya Spetsial'noj Astrofizicheskoj
  Observatorii, \href {https://ui.adsabs.harvard.edu/abs/1973AISAO...5...16K}
  {5, 16}

\bibitem[\protect\citeauthoryear{{Keselman} \& {Nusser}}{{Keselman} \&
  {Nusser}}{2017}]{keselman17}
{Keselman} J.~A.,  {Nusser} A.,  2017, \mn@doi [\mnras] {10.1093/mnras/stx152},
  \href {https://ui.adsabs.harvard.edu/abs/2017MNRAS.467.1915K} {467, 1915}

\bibitem[\protect\citeauthoryear{{Kogut} et~al.,}{{Kogut}
  et~al.}{1993}]{kogut93}
{Kogut} A.,  et~al., 1993, \mn@doi [\apj] {10.1086/173453}, \href
  {https://ui.adsabs.harvard.edu/abs/1993ApJ...419....1K} {419, 1}

\bibitem[\protect\citeauthoryear{{Kopeikin} \& {Makarov}}{{Kopeikin} \&
  {Makarov}}{2006}]{kopeikin06}
{Kopeikin} S.~M.,  {Makarov} V.~V.,  2006, \mn@doi [\aj] {10.1086/500170},
  \href {https://ui.adsabs.harvard.edu/abs/2006AJ....131.1471K} {131, 1471}

\bibitem[\protect\citeauthoryear{{Korzy{\'n}ski} \&
  {Kopi{\'n}ski}}{{Korzy{\'n}ski} \& {Kopi{\'n}ski}}{2018}]{kor18}
{Korzy{\'n}ski} M.,  {Kopi{\'n}ski} J.,  2018, \mn@doi [\jcap]
  {10.1088/1475-7516/2018/03/012}, \href
  {https://ui.adsabs.harvard.edu/abs/2018JCAP...03..012K} {2018, 012}

\bibitem[\protect\citeauthoryear{Kuglin \& Hines}{Kuglin \&
  Hines}{1975}]{kug75}
Kuglin C.~D.,  Hines D.~C.,  1975, Proc. Int. Conference on Cybernetics and
  Society.
IEEE, p. 163–165

\bibitem[\protect\citeauthoryear{{Lange} \& {Page}}{{Lange} \&
  {Page}}{2007}]{lange07}
{Lange} S.,  {Page} L.,  2007, \mn@doi [\apj] {10.1086/523097}, \href
  {https://ui.adsabs.harvard.edu/abs/2007ApJ...671.1075L} {671, 1075}

\bibitem[\protect\citeauthoryear{{Lindegren} et~al.,}{{Lindegren}
  et~al.}{2018}]{lindegren2018}
{Lindegren} L.,  et~al., 2018, \mn@doi [\aap] {10.1051/0004-6361/201832727},
  \href {https://ui.adsabs.harvard.edu/abs/2018A&A...616A...2L} {616, A2}

\bibitem[\protect\citeauthoryear{{Loeb}}{{Loeb}}{1998}]{loeb1998}
{Loeb} A.,  1998, \mn@doi [\apjl] {10.1086/311375}, \href
  {https://ui.adsabs.harvard.edu/abs/1998ApJ...499L.111L} {499, L111}

\bibitem[\protect\citeauthoryear{{Makiya} et~al.,}{{Makiya}
  et~al.}{2016}]{makiya16}
{Makiya} R.,  et~al., 2016, \mn@doi [\pasj] {10.1093/pasj/psw005}, \href
  {https://ui.adsabs.harvard.edu/abs/2016PASJ...68...25M} {68, 25}

\bibitem[\protect\citeauthoryear{{Maller}, {Berlind}, {Blanton}  \&
  {Hogg}}{{Maller} et~al.}{2009}]{maller09}
{Maller} A.~H.,  {Berlind} A.~A.,  {Blanton} M.~R.,   {Hogg} D.~W.,  2009,
  \mn@doi [\apj] {10.1088/0004-637X/691/1/394}, \href
  {https://ui.adsabs.harvard.edu/abs/2009ApJ...691..394M} {691, 394}

\bibitem[\protect\citeauthoryear{{Marshall} et~al.,}{{Marshall}
  et~al.}{2017}]{marshall17}
{Marshall} P.,  et~al., 2017, arXiv e-prints arXiv:1708.04058, \href
  {https://ui.adsabs.harvard.edu/abs/2017arXiv170804058L} {}

\bibitem[\protect\citeauthoryear{{McCrea}}{{McCrea}}{1935}]{mccrea1935}
{McCrea} W.~H.,  1935, \zap, \href
  {https://ui.adsabs.harvard.edu/abs/1935ZA......9..290M} {9, 290}

\bibitem[\protect\citeauthoryear{{M{\'e}ndez}, {Costa}, {Gallart}, {Pedreros},
  {Moyano}  \& {Altmann}}{{M{\'e}ndez} et~al.}{2011}]{mendez11}
{M{\'e}ndez} R.~A.,  {Costa} E.,  {Gallart} C.,  {Pedreros} M.~H.,  {Moyano}
  M.,   {Altmann} M.,  2011, \mn@doi [\aj] {10.1088/0004-6256/142/3/93}, \href
  {https://ui.adsabs.harvard.edu/abs/2011AJ....142...93M} {142, 93}

\bibitem[\protect\citeauthoryear{{Mukherjee}, {Lavaux}, {Bouchet}, {Jasche},
  {Wandelt}, {Nissanke}, {Leclercq}  \& {Hotokezaka}}{{Mukherjee}
  et~al.}{2019}]{mukh19}
{Mukherjee} S.,  {Lavaux} G.,  {Bouchet} F.~R.,  {Jasche} J.,  {Wandelt} B.~D.,
   {Nissanke} S.~M.,  {Leclercq} F.,   {Hotokezaka} K.,  2019, arXiv e-prints
  arXiv:1909.08627, \href
  {https://ui.adsabs.harvard.edu/abs/2019arXiv190908627M} {}

\bibitem[\protect\citeauthoryear{{Nicolaou}, {Lahav}, {Lemos}, {Hartley}  \&
  {Braden}}{{Nicolaou} et~al.}{2019}]{nico19}
{Nicolaou} C.,  {Lahav} O.,  {Lemos} P.,  {Hartley} W.,   {Braden} J.,  2019,
  arXiv e-prints, \href {https://ui.adsabs.harvard.edu/abs/2019arXiv190909609N}
  {p. arXiv:1909.09609}

\bibitem[\protect\citeauthoryear{{Paine}, {Darling}, {Graziani}  \&
  {Courtois}}{{Paine} et~al.}{2020}]{paine20}
{Paine} J.,  {Darling} J.,  {Graziani} R.,   {Courtois} H.~M.,  2020, \mn@doi
  [\apj] {10.3847/1538-4357/ab6f00}, \href
  {https://ui.adsabs.harvard.edu/abs/2020ApJ...890..146P} {890, 146}

\bibitem[\protect\citeauthoryear{{Quercellini}, {Amendola}, {Balbi}, {Cabella}
  \& {Quartin}}{{Quercellini} et~al.}{2012}]{quercellini2012}
{Quercellini} C.,  {Amendola} L.,  {Balbi} A.,  {Cabella} P.,   {Quartin} M.,
  2012, \mn@doi [\physrep] {10.1016/j.physrep.2012.09.002}, \href
  {https://ui.adsabs.harvard.edu/abs/2012PhR...521...95Q} {521, 95}

\bibitem[\protect\citeauthoryear{{Refsdal}}{{Refsdal}}{1966}]{refsdal1966}
{Refsdal} S.,  1966, \mn@doi [\mnras] {10.1093/mnras/132.1.101}, \href
  {https://ui.adsabs.harvard.edu/abs/1966MNRAS.132..101R} {132, 101}

\bibitem[\protect\citeauthoryear{{Riess}}{{Riess}}{2019}]{riess2019}
{Riess} A.~G.,  2019, \mn@doi [Nature Reviews Physics]
  {10.1038/s42254-019-0137-0}, \href
  {https://ui.adsabs.harvard.edu/abs/2019NatRP...2...10R} {2, 10}

\bibitem[\protect\citeauthoryear{{Riess} et~al.,}{{Riess}
  et~al.}{2005}]{riess2005}
{Riess} A.~G.,  et~al., 2005, \mn@doi [\apj] {10.1086/430497}, \href
  {https://ui.adsabs.harvard.edu/abs/2005ApJ...627..579R} {627, 579}

\bibitem[\protect\citeauthoryear{{Riess}, {Casertano}, {Yuan}, {Macri}  \&
  {Scolnic}}{{Riess} et~al.}{2019}]{riess2019b}
{Riess} A.~G.,  {Casertano} S.,  {Yuan} W.,  {Macri} L.~M.,   {Scolnic} D.,
  2019, \mn@doi [\apj] {10.3847/1538-4357/ab1422}, \href
  {https://ui.adsabs.harvard.edu/abs/2019ApJ...876...85R} {876, 85}

\bibitem[\protect\citeauthoryear{{Sandage}}{{Sandage}}{1962}]{sandage62}
{Sandage} A.,  1962, \mn@doi [\apj] {10.1086/147385}, \href
  {https://ui.adsabs.harvard.edu/abs/1962ApJ...136..319S} {136, 319}

\bibitem[\protect\citeauthoryear{Sanderson et~al.,}{Sanderson
  et~al.}{2017}]{sanderson17}
Sanderson R.~E.,  et~al., 2017, Astrometry with the Wide-Field InfraRed Space
  Telescope (\mn@eprint {arXiv} {1712.05420})

\bibitem[\protect\citeauthoryear{{Schlegel} et~al.}{{Schlegel}
  et~al.}{2015}]{schlegel15}
{Schlegel} D.,  et~al., 2015, in APS April Meeting Abstracts. p. Z2.006

\bibitem[\protect\citeauthoryear{{Simard} et~al.,}{{Simard}
  et~al.}{1999}]{simard99}
{Simard} L.,  et~al., 1999, \mn@doi [\apj] {10.1086/307403}, \href
  {https://ui.adsabs.harvard.edu/abs/1999ApJ...519..563S} {519, 563}

\bibitem[\protect\citeauthoryear{{Spergel} et~al.,}{{Spergel}
  et~al.}{2013}]{spergel13}
{Spergel} D.,  et~al., 2013, arXiv e-prints arXiv:1305.5422, \href
  {https://ui.adsabs.harvard.edu/abs/2013arXiv1305.5422S} {}

\bibitem[\protect\citeauthoryear{{Springob} et~al.,}{{Springob}
  et~al.}{2014}]{springob14}
{Springob} C.~M.,  et~al., 2014, \mn@doi [\mnras] {10.1093/mnras/stu1743},
  \href {https://ui.adsabs.harvard.edu/abs/2014MNRAS.445.2677S} {445, 2677}

\bibitem[\protect\citeauthoryear{{Starr} et~al.,}{{Starr}
  et~al.}{2002}]{lsst02}
{Starr} B.~M.,  et~al., 2002, {LSST Instrument Concept}.
Society of Photo-Optical Instrumentation Engineers (SPIE) Conference Series, pp
  228--239, \mn@doi{10.1117/12.457331}

\bibitem[\protect\citeauthoryear{{Tolman}}{{Tolman}}{1934}]{tolman34}
{Tolman} R.~C.,  1934, {Relativity, Thermodynamics, and Cosmology}.
Dover, New York

\bibitem[\protect\citeauthoryear{{Tully}, {Courtois}  \& {Sorce}}{{Tully}
  et~al.}{2016}]{tully16}
{Tully} R.~B.,  {Courtois} H.~M.,   {Sorce} J.~G.,  2016, \mn@doi [\aj]
  {10.3847/0004-6256/152/2/50}, \href
  {https://ui.adsabs.harvard.edu/abs/2016AJ....152...50T} {152, 50}

\bibitem[\protect\citeauthoryear{{Tyson}, {Wittman}, {Hennawi}  \&
  {Spergel}}{{Tyson} et~al.}{2003}]{tyson2003}
{Tyson} J.~A.,  {Wittman} D.~M.,  {Hennawi} J.~F.,   {Spergel} D.~N.,  2003,
  \mn@doi [Nuclear Physics B Proceedings Supplements]
  {10.1016/S0920-5632(03)02073-5}, \href
  {https://ui.adsabs.harvard.edu/abs/2003NuPhS.124...21T} {124, 21}

\bibitem[\protect\citeauthoryear{{Vasiliev}}{{Vasiliev}}{2019}]{vas19}
{Vasiliev} E.,  2019, \mn@doi [\mnras] {10.1093/mnras/stz2100}, \href
  {https://ui.adsabs.harvard.edu/abs/2019MNRAS.489..623V} {489, 623}

\bibitem[\protect\citeauthoryear{Verde, Treu  \& Riess}{Verde
  et~al.}{2019}]{verde2019}
Verde L.,  Treu T.,   Riess A.~G.,  2019, \mn@doi [Nature Astronomy]
  {10.1038/s41550-019-0902-0}, 3, 891–895

\bibitem[\protect\citeauthoryear{Wang, Quan, Liang, Ning, Guo  \& Jiao}{Wang
  et~al.}{2018}]{wang18}
Wang S.,  Quan D.,  Liang X.,  Ning M.,  Guo Y.,   Jiao L.,  2018, ISPRS
  Journal of Photogrammetry and Remote Sensing, 145, 148

\bibitem[\protect\citeauthoryear{{Wang}, {Li}  \& {Cautun}}{{Wang}
  et~al.}{2020}]{wang19}
{Wang} Y.,  {Li} B.,   {Cautun} M.,  2020, \mn@doi [\mnras]
  {10.1093/mnras/staa2136}, \href
  {https://ui.adsabs.harvard.edu/abs/2020MNRAS.497.3451W} {497, 3451}

\bibitem[\protect\citeauthoryear{{Weinberg}}{{Weinberg}}{1972}]{weinberg1972}
{Weinberg} S.,  1972, {Gravitation and Cosmology: Principles and Applications
  of the General Theory of Relativity}.
Wiley

\bibitem[\protect\citeauthoryear{{Willick} \& {Batra}}{{Willick} \&
  {Batra}}{2001}]{willick01}
{Willick} J.~A.,  {Batra} P.,  2001, \mn@doi [\apj] {10.1086/319005}, \href
  {https://ui.adsabs.harvard.edu/abs/2001ApJ...548..564W} {548, 564}

\bibitem[\protect\citeauthoryear{{Yu} \& {Zhu}}{{Yu} \& {Zhu}}{2019}]{yu19}
{Yu} Y.,  {Zhu} H.-M.,  2019, \mn@doi [\apj] {10.3847/1538-4357/ab5580}, \href
  {https://ui.adsabs.harvard.edu/abs/2019ApJ...887..265Y} {887, 265}

\bibitem[\protect\citeauthoryear{Zitova \& Flusser}{Zitova \&
  Flusser}{2003}]{zitova03}
Zitova B.,  Flusser J.,  2003, Image and Vision Computing, 11 edn.
Elsevier, p. 977–1000

\bibitem[\protect\citeauthoryear{{da Silva} et~al.,}{{da Silva}
  et~al.}{2019}]{euclid2019}
{da Silva} R.,  et~al., 2019, {Euclid Near-infrared Imaging Reduction Pipeline:
  Astrometric Calibration, Resampling and Stacking}.
p.~311

\makeatother
\end{thebibliography}

\appendix

\section{Tests of galaxy image registration}

\subsection{Introduction}
As we have seen in the main text, in order to detect parallax shifts of galaxies beyond the local group it will be necessary to make use of more than one astrometric measurement per galaxy. As an example, a Milky-Way
type galaxy at a redshift of $z=0.01$ will extend over $\sim 1$ million NGRST resolution elements. We have assumed the best-case scenario in our analyses in the main text, that the resolution elements combine to make measurements equivalent
to independent sources. In this case, the measurement error is $\propto1/\sqrt{N_{\rm elem}}$ where $N_{\rm elem}$ is the number of resolved elements (see Equation \ref{astroerr} and related discussion). In order for this to
occur, there should be structure in galaxy images, as featureless or smooth galaxies would not have information that could be used to detect image shifts. In this appendix, we carry out some simple  experiments
with observational data in order to get a feeling for how the error on galaxy position scales with  $N_{\rm res}$ for real galaxies.

We caution that carrying out a definitive study of galaxy image registration is far beyond the scope of
this work. For example true tests should include observational and instrumental systematic effects
and errors, and also cover a full representative sample of galaxies. Here we will be working in the
Poisson noise limited regime with relatively low quality images. We will shift galaxy images by a 
small random amount on the plane of the sky and see how well the shift can be 
recovered. Because we are interested in extended images, we will not identify and centroid point sources,
but instead use Fourier space registration techniques.

\subsection{Galaxy data}
Galaxies at a range of distances and with various numbers of resolved elements will be observed by 
both the Rubin and Roman telescopes. We therefore carry out our tests using galaxy images from
both the  ground based SDSS and the HST. The galaxies that we use are primarily spirals and irregulars
(which dominate the datasets that we use), with random orientations, and that are in 
the redshift range that we are interested in, $z < 0.05$.
Elliptical galaxies may lead to larger errors because of their smoother nature, but we leave study of
these and the effect of galaxy type and other properties to future work (which should also take 
into account the relative populations). We choose galaxies of different angular sizes, with half light 
radii spanning from 3 arcsec to 80 arcsec. We also use only one colour, for each galaxy, the SDSS $r$ filter for those images 
and the F606W (red) filter  for HST. The Rubin
and Roman Telescopes will have 6 and 7 filters available respectively, and combining results from
different colour bands should be carried out in those cases to 
improve results.

In the ultimate observational case, the accuracy of image registration is likely to be limited by systematics such as field distortions, charge transfer,
and various other sub pixel effects. Background
sources and  sky noise  (in the case of ground based observations) will also
be important. All the effects are difficult to model, and so are left to
future work. Shot noise from the finite number of photons however is a 
primary source of noise and one which we will concern ourselves with here.
In order for shot noise to be the limiting source in our tests, we must simulate
images which have significantly fewer photons (and shorter effective exposure
times) than the observational data they are derived from. This is in order
that structure caused by noise in the data is not confused with real structures
that can help with image registration.

\begin{figure}
 \includegraphics[width=1.0\columnwidth]{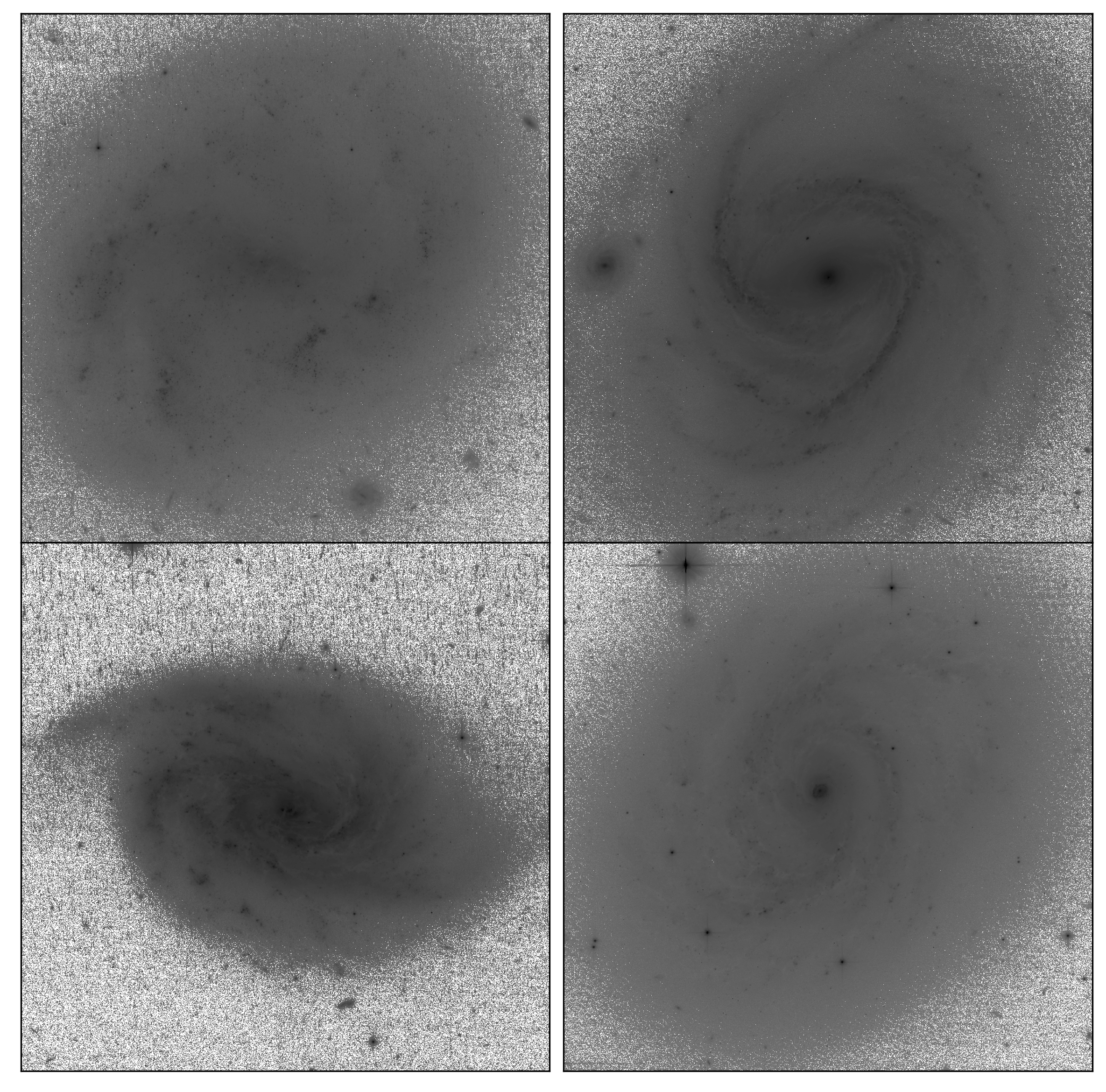}
 \caption{Four of the galaxies with HST ACS imaging used
in our tests of image registration. From top to bottom and left to right the galaxies are NGC1483, NGC1577, NGC1803, and NGC2216. The panels are all 80 arcsecs in vertical and horizontal extent.
}
 \label{hst2x2}
\end{figure}

We use three sets of observational data for our test, two from HST
and one from the SDSS. The first HST set is four galaxies taken from archival data \footnote{ { \tt https://archive.stsci.edu/ } }  (all from HST proposal ID 14840, PI A. Bellini). These were selected to be
 single-orbit imaging of previously unobserved NGC/IC bright galaxies.
 The galaxy IDs and properties
are given in Table \ref{hsttab}. Each galaxy was observed for 696
seconds using the Advanced Camera for 
Surveys Wide Field Channel (ACS/WFC), and we use only one image per galaxy,
which was taken with the F606W filter. Observations
date from 2016-2018. The four HST galaxies used are shown in Figure \ref{hst2x2}. The PSF FHWM 
 of the HST images at this wavelength (used in Equation \ref{astroerr}  is 0.07 arcsec, and the HST pixel size is 0.05 arcsec. We give details of
 their meaasured observational parameters in Table \ref{hsttab}.

The second HST dataset is an archival image of the nearby spiral galaxy NGC2841, taken using the WFC3/UVIS
instrument (HST proposal ID 11360, PI R. O'Connell) The galaxy in this case is significantly closer,
and has $\sim 7$ times more
resolved  elements than the mean for the other HST images used. 

\begin{figure}
 \includegraphics[width=1.0\columnwidth]{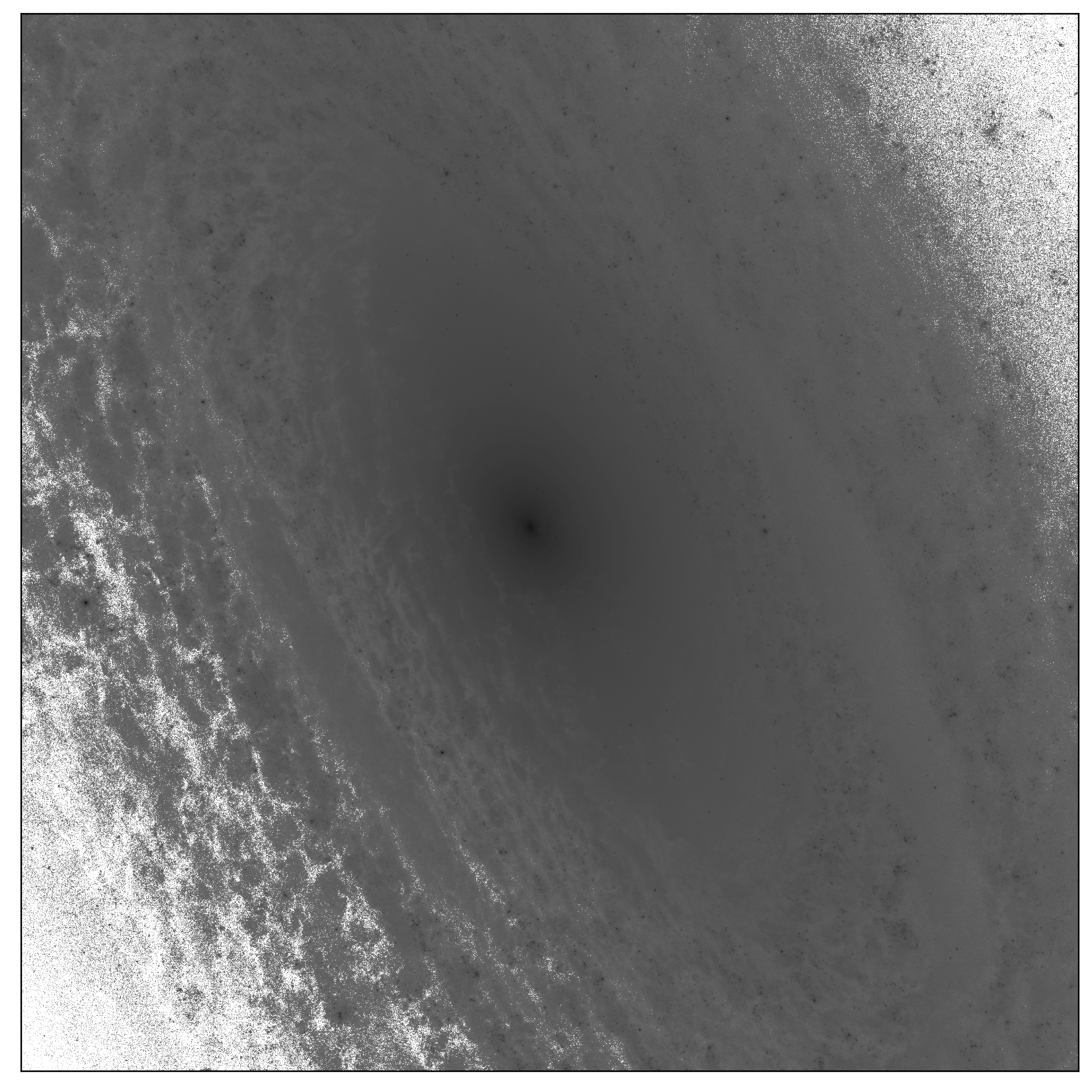}
 \caption{HST/WFC3 image of the region of the nearby spiral galaxy NGC 2841, used for
 our tests of galaxy image registration. The image is a square
 of side length 160 arcsecs.
}
 \label{ngc2841}
\end{figure}

\begin{table}
\begin{tabular}{|l|l|l|l|}
\hline
Galaxy ID & minor axis radius (arcsec) & redshift & $r$ magnitude \\ \hline
NGC1483   & 30.1                       & 0.0039   & 12.26         \\ \hline
NGC1577   & 72.0                       & 0.0312   & 13.0          \\ \hline
NGC1803   & 40.8                       & 0.0136   & 12.54         \\ \hline
NGC2216   & 48.6                       & 0.0091   & 12.34         \\ \hline
NGC2841   & 238.8                      & 0.00190  & 9.88          \\ \hline
\end{tabular}
\caption{Physical properties of
    galaxies with HST imaging used in our
    tests (images are shown in Figures \ref{hst2x2}
    and \ref{ngc2841}). Data is
    from the NASA Extragalactic Database ( {\tt https://ned.ipac.caltech.edu/ }) and the
    CDS SIMBAD database ({\tt https://simbad.u-strasbg.fr } ).}
\label{hsttab}
\end{table}

\begin{figure}
 \includegraphics[width=1.0\columnwidth]{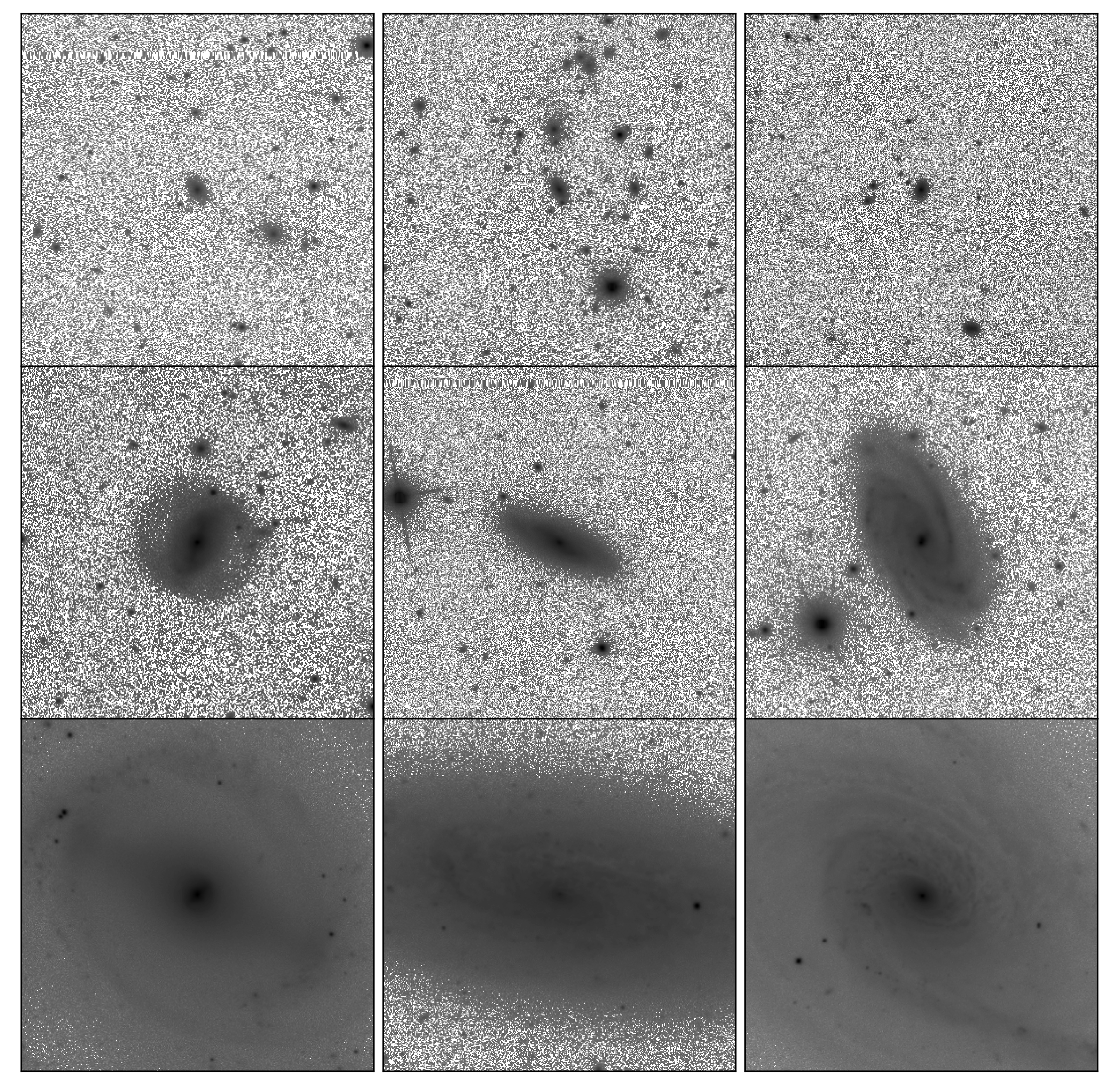}
 \caption{Examples of galaxy images from the NASA Sloan Atlas used for our tests of image registration. The top row shows galaxies with Petrosian radius between 3 and 5 arcsecs, middle row
 10-30 arcsecs, and bottom row 70-80 arcsec. All images are 160 arcsecs in diameter. 
}
 \label{nsa3x3}
\end{figure}

To carry out tests with galaxies that have few resolution elements,
our third dataset consists of galaxy images taken from the
ground based SDSS telescope. The NASA-Sloan atlas\footnote{ {\rm \tt http://nsatlas.org } } is a 
 catalogue of images (see e.g., \cite{maller09} and parameters of local galaxies derived from SDSS imaging, and with the addition of Galaxy Evolution Explorer (GALEX) data for the ultraviolet part of the spectrum. We do not
use the UV data, but as with the HST data above we restrict ourselves
to a single filter image per galaxy (in the SDSS $r$ band). We 
 choose 12 spiral or irregular galaxies randomly in each of three bins of Petrosian radius, 3-5 arcsec, 10-30 arcsec and 70-80 arcsec.  Some example galaxies in each of these three bins are shown in Figure \ref{nsa3x3}. In the case of the NSA galaxies, the seeing will determine the size of the resolution
 elements. The mean SDSS $r$ band imaging seeing is 1.32 arcsec FWHM, and for simplicity we use this value for all NSA images. The SDSS pixel size is 0.40 arcsecs.

\subsection{Method}

For each galaxy image, we first select a circular 
region within two Petrosian radii of the galaxy centroid
in the case of the NSA data, and two times the minor axis radius in the case of the large
HST galaxies. The number of resolution elements in the galaxy is then computed using Equation \ref{nelemeq}.
In the case of NGC 2841, we split this region into separate square chunks of side length 400 pixels,
in order to test the effect of using mosaic images. Our results for this galaxy are averaged over all chunks.

We shift each galaxy image by a distance in the $x$ and $y$ directions drawn  from uniform
random distributions between -10 and +10 pixels. The image shift is carried out in 
Fourier space, using the Python routine {\tt scipy.ndimage.fourier\textunderscore shift }. We restrict ourselves in 
these tests to rigid translations, leaving rotations and deformations to future work.

The next step is to add noise to each of the images (shifted and unshifted). We do this by Poisson sampling the image with discrete photons. In order to make sure that we are in the Poisson limited regime, we simulate  short exposures, using an effective exposure time of one second for the NSA images (compared to  the SDSS single exposure time of 54 seconds). This leads to a mean number of photons per resolution element of 200. To be consistent and allow comparison across the different galaxy datasets and  angular sizes, we also scale the effective exposure time of the HST images so that the mean number of photons per resolution element is also 200.

At this point, we have two noisy images of the same galaxy and would like to estimate the image
shift. In our case, we have introduced a shift which is a rigid translation.
There are a number of methods which can be employed to determine the shift, and the 
literature on this topic is growing (e.g., \citealt{wang18} ). We use a 
simple Fourier
method, based on publicly available software, the {\tt astropy} image registration 
package\footnote{\tt https://image-registration.readthedocs.io/en/\\
latest/image\_registration.html}. The specific routine
we use is {\tt chi2\_shift }. This is based on the single step Discrete Fourier Transform 
algorithm from \cite{guizar08}. The algorithm
uses the entire image to compute the upsampled cross-correlation in Fourier
space in a small neighborhood around its peak. This allows rapid subpixel registration, but has not been shown to be an optimal method.

We carry out the process 100 times for each galaxy image, with different photon sampling
and image shifts each time, and then compute the {\it rms} error on the detected shift. 

\subsection{Results}
\begin{figure}
 \includegraphics[width=1.0\columnwidth]{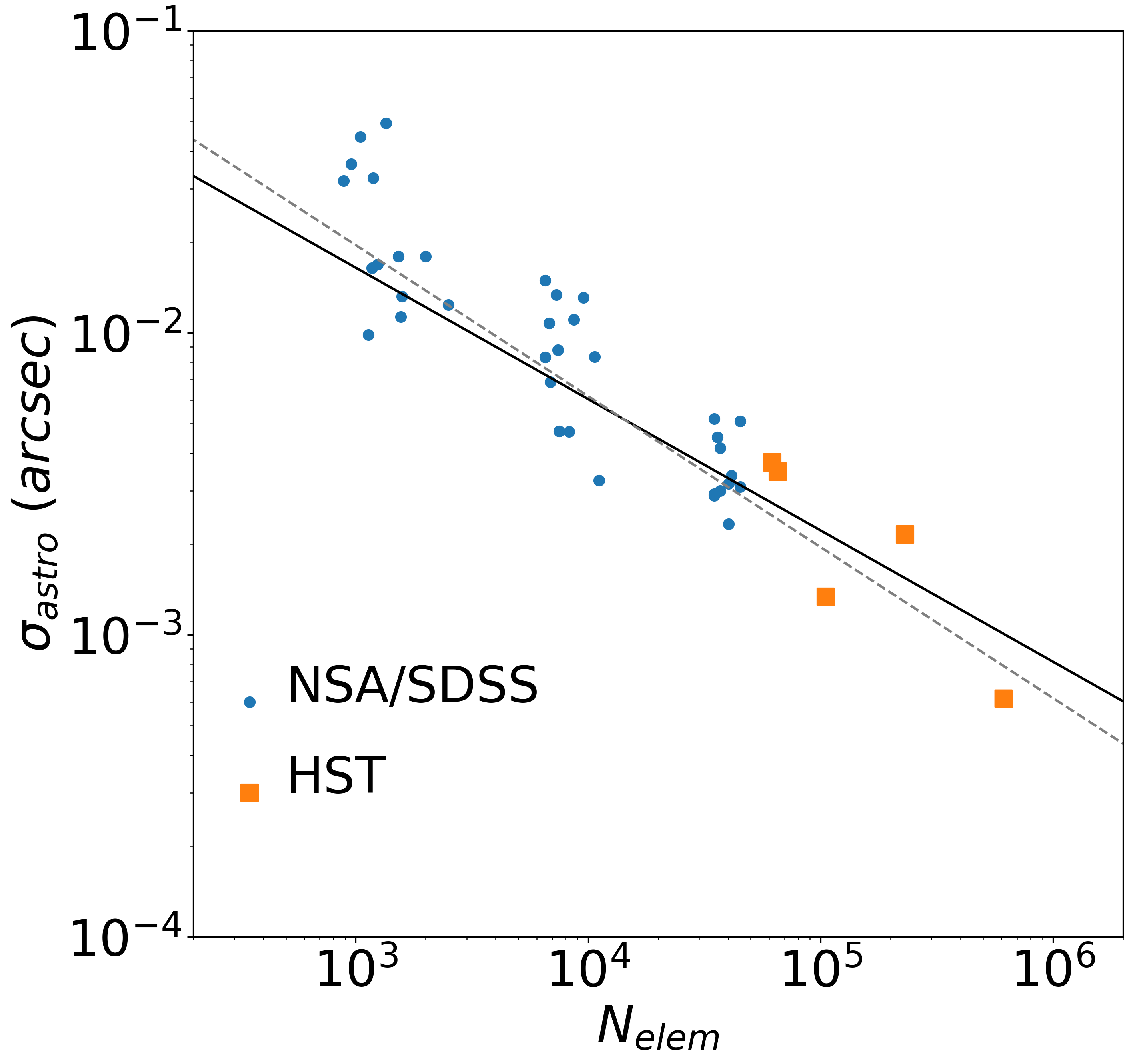}
 \caption{The {\it rms} error, $\sigma_{\rm astro}$ on measurements of galaxy image shifts using
 Fourier space image registration techniques plotted as a function of the number of resolved elements in each individual galaxy, $N_{\rm elem}$ 
 (computed using Equation \ref{nelemeq}). We show results for the
 HST images as orange squares and the ground based data from SDSS as
 blue points. The solid line is a best fit to all points, and has
 a slope $\sigma_{\rm astro} \propto N_{\rm elem}^{-0.440\pm0.025}$.
 The dashed line is the ideal scaling,  $\sigma_{\rm astro} \propto N_{\rm elem}^{-0.5}$.
}
 \label{sigmavsres}
\end{figure}

In Figure \ref{sigmavsres} we show the {\it rms} error on the measured translational shifts, $\sigma_{\rm astro}$ as points for each galaxy, as a function of $N_{\rm elem}$, the number of resolution elements. 
In order to plot both HST and NSA galaxies on the same graph, we have converted the NSA results
from arcsec into pixels and then into arcsec at the resolution of HST. The mean number of photons per resolution element was also the same for all galaxies, as described above.

We can see that the accuracy of shift measurement increases with the number of resolution elements, 
although there is significant scatter. We have carried out a least squares fit of the best fit power law (with residuals in log $\sigma_{\rm astro}$, using the Python routine {\tt scipy.optimize.curve\_fit}), and find the slope to be $-0.440\pm0.025$. We show this best fit line on Figure \ref{sigmavsres} as a solid line. 
We also show in Figure \ref{sigmavsres} a dashed line corresponding to the case where 
$\sigma_{\rm astro} \propto N_{\rm elem}^{-0.5}$, which we have assumed in our analyses in the main text. We can see that this is somewhat steeper than the fit relationship, but only at the $2.4 \sigma$ level. Real galaxies do not therefore appear to behave very much differently than our assumptions
in the text, at least as far as these simple tests show.

\subsection{Discussion}

We have carried out some simple tests with observational data to find the accuracy with which galaxy image shifts can be detected. We have found that
the positions of
larger galaxies, spanned by more resolution elements can indeed be found more
accurately than smaller galaxies. The absolute value of the slope of the relationship between
the number of elements and {\it rms} measurement error is slightly less than the
the ideal value of 0.5, which was used in our predictions in the main text. We have decided not to change the fiducial value in the text, because the tests here
are merely illustrative, and many questions remain to be answered (see below). 

Our tests have been limited to the shot noise dominated regime, with high noise levels, due to the necessity to avoid conflating noise features in the test data
with real structure. This is
obviously the easiest regime to simulate, but has meant that we have 
not been able
to test the extremely small image shifts that will be present in Rubin
and Roman data. These shifts will be far below our test shifts (for example,
the point source measurement error for NGRST is $\sim2.5\times10^{-5}$ arcsecs). 
Also, there will be systematic errors that enter,  including field distortions, colour/filter effects, charge transfer effects and more. A more sophisticated treatment will need to deal with these.

Another important aspect that we have not touched are the magnitudes of the errors themselves. We have restricted ourselves to relative comparisons between different sized galaxies under the same conditions. There are two reasons for this, the first being that different techniques for image registration of extended objects at the sub pixel level could perform much better or worse than others. For example, \cite{hrazd20} have shown how recently developed 
iterative methods can offer
substantial improvements on some prior methods. The second is that although it would obviously be useful to compare the accuracy to that of centroiding point sources (particularly since
the point source {\it rms} errors are what is quoted for the telescopes of interest), 
simulating point sources in this context is non-trivial. For point source tests, 
one would need to model a realistic background, detection of sources themselves,
and determine how non-detections would affect the errors. We leave these and other  complexities to future work.

In the main text, we have discussed that a reference frame will need to be set up for cosmic
parallax measurements based on extremely distant galaxies (for example
at distances greater than 1 Gpc from the observer, where parallaxes will
not be measurable). We have seen that the number density of distant galaxies
is expected to be high enough that the errors from setting the frame will
be very subdominant. We have not addressed this issue in our tests, but have instead assumed measurements of galaxy image shifts with respect to a constant
reference frame defined by pixel axes. We have also only addressed rigid translation of images. In reality, there will be there will be
other rigid transformations (rotations), as well as affine (shears for example), homographies, or more complex large deformations. Image registration 
techniques used will have to deal with these. There is therefore much that 
has been left to future work.

\end{document}